\documentclass[]{emulateapj}
\usepackage{apjfonts}
\usepackage{graphicx}
\usepackage{textcomp}

\def\lsim{\lower.5ex\hbox{$\; \buildrel < \over \sim \;$}}
\def\gsim{\lower.5ex\hbox{$\; \buildrel > \over \sim \;$}}
\newcommand{\msun}{$M_\odot$}

\shorttitle{The origin of Beryllium-10 in the protosolar nebula}

\shortauthors{Tatischeff, Duprat, \& de S\'er\'eville}

\begin{document}


\title{Light element nucleosynthesis in a molecular cloud interacting with a supernova remnant and the origin of Beryllium-10 in the protosolar nebula}

\author{Vincent Tatischeff}
\author{Jean Duprat}
\affil{Centre de Sciences Nucl\'eaires et de Sciences de la Mati\`ere, 
IN2P3-CNRS and Univ Paris-Sud, F-91405 Orsay Cedex, France}
\email{Vincent.Tatischeff@csnsm.in2p3.fr}

\and

\author{Nicolas de S\'er\'eville}
\affil{Institut de Physique Nucl\'eaire d'Orsay, 
IN2P3-CNRS and Univ Paris-Sud, F-91405 Orsay Cedex, France}

\begin{abstract}
The presence of short-lived radionuclides ($t_{1/2}< 10$~Myr) in the early solar system provides important information about the astrophysical environment in which the solar system formed. The discovery of now extinct $^{10}$Be ($t_{1/2}=1.4$~Myr) in calcium-aluminum-rich inclusions (CAIs) with Fractionation and Unidentified Nuclear isotope anomalies (FUN-CAIs) suggests that a baseline concentration of $^{10}$Be in the early solar system was inherited from the protosolar molecular cloud. In this paper, we investigate various astrophysical contexts for the nonthermal nucleosynthesis of $^{10}$Be by cosmic-ray-induced reactions. We first show that the $^{10}$Be recorded in FUN-CAIs cannot have been produced in situ by irradiation of the FUN-CAIs themselves. We then show that trapping of Galactic cosmic rays (GCRs) in the collapsing presolar cloud core induced a negligible $^{10}{\rm Be}$ contamination of the protosolar nebula, the inferred  $^{10}{\rm Be} / ^{9}{\rm Be}$ ratio being at least 40 times lower than that recorded in FUN-CAIs ($^{10}{\rm Be} / ^{9}{\rm Be} \sim 3 \times 10^{-4}$). Irradiation of the presolar molecular cloud by background GCRs produced a steady-state $^{10}{\rm Be} / ^{9}{\rm Be}$ ratio  $\lsim 1.3\times10^{-4}$ at the time of the solar system formation, which suggests that the presolar cloud was irradiated by an additional source of CRs. Considering a detailed model for CR acceleration in a supernova remnant (SNR), we find that the $^{10}$Be abundance recorded in FUN-CAIs can be explained within two alternative scenarios: (i) the irradiation of a giant molecular cloud by CRs produced by $\gsim 50$ supernovae exploding in a superbubble of hot gas generated by a large star cluster of at least 20,000 members and (ii) the irradiation of the presolar molecular cloud by freshly accelerated cosmic rays escaped from an isolated SNR at the end of the Sedov-Taylor phase. In the second picture, the  SNR resulted from the explosion of a massive star that ran away from its parent OB association, expanded during most of its adiabatic phase in an intercloud medium of density of about $1$~H-atom~cm$^{-3}$, and eventually interacted with the presolar molecular cloud only during the radiative stage. This model naturally provides an explanation for the injection of other short-lived radionuclides of stellar origin into the cold presolar molecular cloud ($^{26}$Al, $^{41}$Ca and $^{36}$Cl) and is in agreement with the solar system originating from the collapse of a molecular cloud shocked by a supernova blast wave. 
\end{abstract}


\section{Introduction}

Several short-lived radionuclides (SLRs) with half-lives $t_{1/2}< 10$~Myr were present in the protoplanetary disk at the time of the solar system formation, including $^{10}$Be, $^{26}$Al, $^{36}$Cl, $^{41}$Ca, $^{53}$Mn, $^{60}$Fe, $^{107}$Pd, and $^{182}$Hf \citep[see][for a  review]{dau11}. These SLRs have various origins. Some, such as $^{53}$Mn ($t_{1/2}=3.7$~Myr), $^{107}$Pd ($t_{1/2}=6.5$~Myr), and $^{182}$Hf ($t_{1/2}=8.9$~Myr) had initial abundances in the early solar system consistent with the levels expected from the long-term chemical evolution of the Galaxy \citep{mey00}. Others, such as $^{26}$Al ($t_{1/2}=0.72$~Myr), were most probably synthesized in a nearby stellar source just before or during the birth of the solar system \citep[see, e.g.,][]{vil09,tat10}. Various stellar sources of $^{26}$Al have been proposed in the literature since the discovery of excess $^{26}$Mg (the decay product of $^{26}$Al) in calcium-aluminum-rich inclusions (CAIs) from the Allende meteorite \citep{lee76}: a core-collapse supernova \citep[SN;][]{cam77,pan12}, an asymptotic giant branch (AGB) or super-AGB star \citep{was94,lug12}, a Wolf-Rayet star \citep{arn97,tat10}, and more recently a massive star (of mass $M > 30~M_\odot$) on the main sequence \citep{gou12}. 

Since the first evidence of $^{60}$Ni excesses in Allende CAIs \citep{bir88}, the initial value of the $^{60}{\rm Fe} / ^{56}{\rm Fe} $ ratio in the early solar system remains debated. In-situ measurements suggest a high initial abundance of $^{60}$Fe ($t_{1/2}=2.6$~Myr) in chondrites, at a level of $^{60}{\rm Fe} / ^{56}{\rm Fe} \sim  10^{-7}$ \citep[see, e.g.,][]{tac06,mis10}, while bulk measurements in several types of meteorites and in sub-components (e.g. chondrules) exhibit much lower ratios, typically of the order of $\sim 10^{-8}$. Whereas the latter value is compatible with the average interstellar medium (ISM) composition at the time of the solar system formation \citep{tan12}, the high value may reveal a possible contamination of the early solar system by material from a nearby SN \citep{was98}. However, caution is required in interpreting these data. The inferred initial $^{60}{\rm Fe} / ^{56}{\rm Fe}$ ratio can be biased by data reduction  \citep{Ogliore2011,Telus2012} and the possible redistribution of Fe (after the $^{60}$Fe decay) can result in a substantially higher (or lower) apparent $^{60}{\rm Fe} / ^{56}{\rm Fe} $ ratio \citep{Telus2014}. As a result, in the current level of knowledge, the $^{60}$Fe data do not provide clear evidence that the solar nebula was substantially polluted by a SN responsible for triggering the gravitational collapse of the protosolar molecular cloud. 

A key question raised by these observations is to understand to what extend the birth of the solar system occurred in generic conditions or require a specific astrophysical context involving an unusual sequence of stellar events. The substantial contamination of the protosolar nebula by specific stellar nucleosynthetic products carried by winds and/or SN ejecta necessarily requires ad hoc assumptions on the immediate neighborhood of the solar system at the time of its formation. However, in the past years, attempts were made to explain the presence of most SLRs in a global generic approach taking into account the astronomical observation of star forming regions \citep{gou12,you14,sah14}. In the model proposed by \citet{gou12}, the early solar system $^{60}$Fe budget comes from a first generation of stars, whereas $^{26}$Al is injected later by the winds of a specific massive star of second generation located at close distance from the gas shell in which the solar system was born \citep[see also][]{gou14}. More recently, \citet{you14} showed that the abundance of most SRLs can be explained within a generic Galactic evolution model taking into account the capture of massive-star winds in dense molecular clouds. 
Within such a context, $^{10}$Be ($t_{1/2}=1.4$~Myr) is of uttermost interest since it is not produced by stellar nucleosynthesis, but only by accelerated particle induced reactions \citep{ree94}. Therefore, it was not taken into account in the approaches mentioned above, but was assumed to arise from another context. In this paper, we investigate different non-thermal nucleosynthesis scenarios for $^{10}$Be that shed light on the astrophysical context of the solar system formation.

This SLR was present in various CAIs \citep[e.g.][]{mck00,cha06,wie12} and refractory hibonites \citep[e.g.][]{mar02,liu09} with an initial ratio $^{10}{\rm Be} / ^{9}{\rm Be}$ ranging from $\sim 3 \times 10^{-4}$ to $\sim 9 \times 10^{-4}$ \citep[see][and references therein]{sri13}. Recently, \citet{gou13} reported the detection of a much higher initial concentration of $^{10}$Be in one CAI from the chondrite Isheyevo: $^{10}{\rm Be} / ^{9}{\rm Be} = (10.4 \pm 1.6)\times 10^{-3}$. Such a large variation of the $^{10}{\rm Be} / ^{9}{\rm Be}$ ratio can be explained if substantial amounts of $^{10}$Be were synthesized within the early solar system by solar energetic particle irradiation of gas and/or dust located at the inner edge of the protoplanetary disk \citep{gou06,gou13,ley03,liu09}. In this model, the variability of the $^{10}$Be abundance is a result of the large range of irradiation dose received by target material during flaring events. By analogy with the present day implantation of $^{10}$Be in lunar materials, \cite{bri10} proposed an alternative model in which $^{10}$Be is synthesized by large flares in the atmosphere of the young Sun, and then incorporated into the solar wind and implanted into refractory solids at the edge of the protoplanetary disk. 

\citet{wie12} recently identified the past presence of live $^{10}$Be in two CAIs with Fractionation and Unidentified Nuclear isotope anomalies (FUN-CAIs), with an initial ratio $^{10}{\rm Be} / ^{9}{\rm Be} \sim 3 \times 10^{-4}$. FUN-CAIs are rare refractory igneous objects that formed very early in the history of the solar system, most likely prior to the condensation of canonical CAIs, and had minor interaction with any solar gas \citep[see][]{thr08}. Therefore, \citet{wie12} suggested that the $^{10}{\rm Be} / ^{9}{\rm Be}$ ratio recorded by FUN-CAIs represents a baseline level present in presolar material inherited from the protosolar molecular cloud. \citet{wie12} further suggested that the $^{10}$Be enrichment of the protosolar molecular cloud results from the trapping of Galactic cosmic rays (GCRs) in the collapsing cloud core \citep{des04}. 

Gamma-ray astronomy has recently provided direct evidence that CR protons are accelerated in SN remnants (SNRs), where they produce neutral pions by interacting with  ambient thermal gas \citep{ack13}. In some objects, part of the pionic gamma-ray emission arises from outside the remnant and is most likely due to CRs that escaped the acceleration site and now interact with nearby molecular clouds \citep{abd10,uch12}. Inspired by these gamma-ray observations --~hadronic gamma-ray emission is necessarily accompanied by spallation nucleosynthesis of light elements~-- we investigate the possibility that the $^{10}$Be abundance recorded by FUN-CAIs was produced by CRs accelerated in a nearby SNR. 

The main goal of this paper is to use the meteoritic data on protosolar $^{10}$Be to study the assumption that the birth of the solar system was triggered by a SN shock. In Section~2, we develop a detailed model for the production of hadronic CRs at the blast wave of a SNR and the transport of these energetic particles in the postshock plasma. We then use the derived particle distribution functions to calculate light element nucleosynthesis as a function of the ambient medium density. In Section~3, we first show that $^{10}$Be found in FUN-CAIs was most probably not produced by nonthermal nucleosynthesis within the early solar system (Sect.~3.1) and then study various models of $^{10}$Be synthesis in the protosolar molecular cloud (Sect.~3.2). In the following of this Section, based on recent results on the abundance of $^{10}$Be nuclei in current-epoch GCRs, we first show that the GCR-trapping model put forward by \citet{des04} cannot account for $^{10}{\rm Be} / ^{9}{\rm Be} \sim 3 \times 10^{-4}$ at the time of the solar system formation. We then evaluate the amount of protosolar $^{10}$Be produced by irradiation of the presolar molecular cloud by background GCRs and highlight the need for an additional source of CRs producing nonthermal nucleosynthesis in the presolar gas. We then use the $^{10}$Be production rate derived in Sect. 2 to explore two models of SN-molecular cloud interaction: one where the radioactive nuclei are produced inside the SNR by trapped CRs interacting with shocked molecular gas, and another one where $^{10}$Be is produced in a molecular cloud irradiated by freshly accelerated CRs that escaped from the remnant during the radiative stage of its evolution. In the continuation of the latter model, we also consider a scenario where the $^{10}$Be-producing CRs are accelerated in several SNe exploding in a superbubble of hot gas resulting from the activity of massive stars in an OB association. The results of the various models are discussed in Section~4 and a conclusion is finally given in Section~5.

\section{Model of light element nucleosynthesis by CR spallation in an SNR}

\subsection{Rate of cosmic-ray production at SN blast wave}

Both gamma-ray emission from SNRs \citep[e.g.,][]{gio12} and models for the origin of the CRs detected near Earth \citep[e.g.,][]{ptu10} show that a sizeable fraction of the kinetic motion of an expanding SN shell gets converted into accelerated particles, most likely by the diffusive shock acceleration process \citep[][and references therein]{cap12}. In a young SNR, typically 50\% of the ram kinetic energy flux processed by the blast wave goes into CRs at any instant \citep{ell11}. About 20\% of this energy flux is given to particles that continuously escape upstream into the ISM, while the remaining CRs carrying 30\% of the available kinetic energy are trapped within the SNR until it becomes radiative \citep{cap10,ell11}. In the present work, we consider only the production of light elements by trapped CRs. We assume that the total kinetic power acquired at any instant by these particles at the forward shock front is given by
\begin{equation}  
\dot{W}_{\rm CR} = f_{\rm CR}\dot{W}_s =f_{\rm CR} \times \frac{1}{2} \rho _{\rm CSM} V_{s}^{3} \times 4\pi R_{s}^{2}~,      
\label{eq1}
\end{equation} 
with $f_{\rm CR}=30$\%. Here, $\rho _{\rm CSM}$ is the density of the circumstellar medium (CSM) of the SN, $R_s$ and $V_s=dR_s/dt$ being the forward shock radius and velocity, respectively. Assuming power-law density profiles for both the outer SN ejecta, $\rho_{\rm ej} \propto t^{n-3} R^{-n}$ (with $n>5$), and the CSM, $\rho _{\rm CSM} \propto R^{-s}$ (with $s<3$), we have from the self-similar, thin-shell approximation \citep{che82} that during the initial free expansion phase of the SNR: $R_s \propto t^{(n-3)/(n-s)}$. Equation~(\ref{eq1}) then gives $\dot{W}_{\rm CR} \propto t^m$ with $m=(2n+6s-ns-15)/(n-s)$. If $\rho _{\rm CSM}$ is constant ($s=0$), then $m=2-15/n$; but if $\rho _{\rm CSM}$ drops as $R^{-2}$, which corresponds to the case where the blast wave initially expands in the winds of the progenitor star, then $m=-3/(n-2)$. The power-law index of the outer SN ejecta being typically in the range $8<n<12$ \citep[see, e.g.,][]{mat99}, we obtain $0.125<m<0.75$ for $s=0$ and $-0.5<m<-0.3$ for $s=2$. Thus, in an SNR resulting from the explosion of a massive star that experienced strong mass loss at the end of its life, the CR power $\dot{W}_{\rm CR}$ is expected to be maximum just after the outburst. 

For simplicity, we neglect in this work the light element nucleosynthesis during the free expansion phase and further assume that the SNR expands into a CSM of constant density. This will allow us to obtain a lower limit on $^{10}$Be production in an SNR, which will be independent of the SN type and the wind mass loss of the progenitor star. We note, however, that in remnants of massive star explosions, the production of $^{10}$Be during the early stage of interaction of the SN ejecta with the progenitor wind might be significant. 

The free expansion phase ends when the mass of interstellar matter swept up and collected by the forward shock becomes comparable to the mass of the SN ejecta, $M_{\rm ej}$, which occurs at the time after explosion \citep{tru99}
\begin{equation} 
t_{\rm ST} \approx \left(1400 {\rm \; yr}\right)\left(\frac{M_{\rm ej} }{10~M_\odot} \right)^{5/6} \left(\frac{E_{\rm SN} }{10^{51} {\rm \; erg}} \right)^{-1/2} \left(\frac{n_{\rm H} }{1{\rm \; cm}^{{\rm -3}} } \right)^{-1/3}.    
\label{eq2} 
\end{equation} 
Here, $E_{\rm SN}$ is the total kinetic energy of the SN outburst and $n_{\rm H}$ the H number density in the CSM. During the subsequent adiabatic Sedov-Taylor stage, the forward shock radius evolves as
\begin{equation} 
R_{s} =\left(12.5{\rm \; pc}\right)\left(\frac{E_{\rm SN} }{10^{51} {\rm \; erg}} \right)^{1/5} \left(\frac{n_{\rm H} }{1{\rm \; cm}^{{\rm -3}} } \right)^{-1/5} \left(\frac{t}{10^{4} {\rm \; yr}} \right)^{2/5},     
\label{eq3} 
\end{equation} 
such that $\dot{W}_{\rm CR} \propto t^{-1}$ (see Eq.~\ref{eq1}). The transition from the Sedov-Taylor stage to the radiative pressure-driven snowplow phase occurs at the time \citep{blo98}
\begin{equation} 
t_{\rm rad} \approx \left(2.9\times 10^4 {\rm \; yr}\right)\left(\frac{E_{\rm SN} }{10^{51} {\rm \; erg}} \right)^{4/17} \left(\frac{n_{\rm H} }{1{\rm \; cm}^{{\rm -3}} } \right)^{-9/17}.    
\label{eq4} 
\end{equation} 
In the radiative phase, the thermal gas in the shell of the swept-up material gradually recombines, which has the effects of terminating the process of particle acceleration and allowing the CRs previously accelerated to escape into the ISM. 

Considering only the CRs accelerated during the Sedov-Taylor stage, the temporal evolution of the energy density of these particles at the forward shock position can be written as \citep[see][]{par99b}
\begin{equation} 
\epsilon _{\rm CR}(R_s,t) =\frac{\dot{W}_{\rm CR}}{4\pi R_s^2 V_s}  \approx \frac{f_{\rm CR} E_{\rm SN} }{4\pi R_s^2 V_s t} .     
\label{eq5} 
\end{equation} 
By definition, the CR energy density is also given by
\begin{equation}  
\epsilon _{\rm CR}(R_s,t) =\int _{p_{\min } }^{p_{\max } }4\pi p^2 f(p,t)E(p)dp~,     
\label{eq6}
\end{equation} 
where $p$ and $E$ are the particle momentum and kinetic energy, respectively, and
\begin{equation}  
f(p,t)=f_0(t) \left(\frac{p}{mc} \right)^{-s_p}~     
\label{eq7}
\end{equation} 
is the CR phase-space distribution expected from the diffusive shock acceleration theory ($m$ is the particle mass and $c$ the speed of light). Typical limits of the CR momentum during the Sedov-Taylor stage are $p_{\min}\sim10^{-3} mc$ and $p_{\max}\sim 10^6 mc$ \citep[e.g.][]{cap12}. The differential number density of CRs per unit energy interval --~expressed for example in number of particles cm$^{-3}$~(MeV/nucleon)$^{-1}$~-- is related to the phase-space distribution by
\begin{eqnarray}  
n(E,R_{s},t)&&=4\pi p^{2} f(p,t)\frac{dp}{dE} \nonumber \\
            &&=4\pi f_0(t) \left(\frac{p}{mc} \right)^{1-s_p } m^{2} c\left(\frac{E}{mc^{2} } +1\right). 
\label{eq8}
\end{eqnarray}

Recent gamma-ray observations of Galactic SNRs show that the energy spectrum of relativistic CRs accelerated in these objects is proportional to  $E^{-s_E}$ with $s_E$ in the range $2.2 - 2.4$, which is steeper than the $E^{-2}$ dependence predicted by the test-particle model of first-order Fermi acceleration \citep[][and references therein]{cap11}. A CR source spectrum as steep as $E^{-2.2}-E^{-2.4}$ is also needed to explain the slope of the CR flux observed near Earth ($\propto E^{-2.75}$). The steepness of the CR source spectrum can be explained by the Alfv\'enic drift of self-generated plasma waves in the precursor regions of SN shocks (\citealt{cap11,cap12}; see also \citealt{bel78}). To account for this effect, we adopt for the slope of the phase-space distribution $s_p=s_E+2=4.3 \pm 0.1$. 

The integral in equation~(\ref{eq6}) can be readily calculated by using as a first approximation $E(p)=p^2/2m$ for $p\leq mc$ and $E(p)=pc$ for $p>pc$ \citep[see also][]{dru89}:
\begin{equation}  
\epsilon _{\rm CR}(R_s,t) =4\pi f_0(t) m^{4} c^{5} \Im~,   
\label{eq9}
\end{equation} 
with 
\begin{equation}  
\begin{array}{l} {\Im =\left[\frac{1-\left(p_{\min } /mc\right)^{5-s_p } }{5-s_p } +\frac{\left(p_{\max } /mc\right)^{4-s_p } -1}{4-s_p } \right]{\rm ~for~}s_p \ne {\rm 4\; and\; 5}} \\ 
{\Im =1-\left(\frac{p_{\min } }{mc} \right)+\ln \left(\frac{p_{\max } }{mc} \right){\rm ~~~~~~~~~~~~~~~~~~~for~}s_p ={\rm 4\; }} \\ 
{\Im =1-\ln \left(\frac{p_{\min } }{mc} \right)-\left(\frac{mc}{p_{\max } } \right){\rm ~~~~~~~~~~~~~~~~~~for~}s_p ={\rm 5\;. }} \end{array}
\label{eq10}
\end{equation} 
For $4\leq s_p \leq 5$, the calculation of $\epsilon _{\rm CR}$ depends little on the values of $p_{\min}$ and $p_{\max}$. Moreover, the approximation of $E(p)$ used for this calculation leads to a negligible error in the result. 

By expressing in equation~(\ref{eq8}) the time-dependent normalization factor $f_0(t)$ from equations~(\ref{eq5}) and (\ref{eq9}), we obtain:
\begin{equation}  
n(E,R_{s} ,t)=\frac{Q_{s}(E,t)}{4\pi R_{s}^{2} V_{s}}~,
\label{eq11}
\end{equation} 
with
\begin{equation}  
Q_{s}(E,t)=\frac{f_{\rm CR} E_{\rm SN}}{t} \frac{1}{\Im m^{2} c^{4} } \left(\frac{p}{mc} \right)^{1-s_p} \left(\frac{E}{mc^{2} } +1\right)~. 
\label{eq12}
\end{equation} 
The latter quantity is the differential injection rate of CRs at the forward shock front (number of particles s$^{-1}$~(MeV/nucleon)$^{-1}$). Its expression in the form of equation~(\ref{eq12}) is a generalization of equation~(16) in \citet{par99b}, valid for any value of $s_p$. Using $Q_{s}(E,t)$ to describe the CR injection rate presupposes that these particles are instantaneously produced at the shock position whatever their energy. This approximtaion is sufficient for the present model, because the light elements are mainly produced by spallation induced by non-relativistic CRs, and the acceleration time of these particles at SN shocks is much shorter than the dynamic timescale of SNRs.  

\subsection{Cosmic-ray transport in the postshock plasma}

Because nonthermal particles in SNRs are thought to be strongly coupled to the thermal plasma through the self-generated magnetic turbulence, we assume that CRs are transported away from the shock front by simple advection with the downstream thermal fluid (see, e.g., \citealt{par06} for a study of high-energy CR electron transport in SNRs). The differential number density of CRs of type $j$ (protons, $\alpha$  particles and heavier nuclei) at time $t$ and position $R<R_s(t)$, denoted $n_j(E,R,t)$, then results from the advection of the corresponding nonthermal population generated at the forward shock front at time $t_i<t$: $n_j(E_i,R_s(t_i),t_i)$ (eqs.~\ref{eq11} and \ref{eq12})\footnote{Note that $R>R_s(t_i)$ since the SNR is expanding over time.}. The correspondence between $R$, $t$ and $t_i$ is obtained by the numerical solution of the equation
\begin{equation}
R(t)=R_s(t_i)+\int_{t_i}^{t} V[R(t')]dt'~,     
\label{eq13} 
\end{equation} 
where $V[R(t')]$ is the bulk velocity of the downstream thermal fluid. We estimate the latter from the self-similar Sedov solution \citep[see][]{che83}, neglecting the influence of the CR pressure on the shock structure. 

\begin{figure}
\centering
\includegraphics[width=7.2cm]{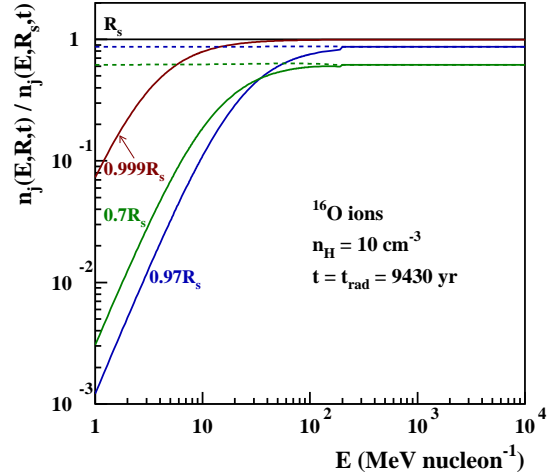}
\caption{Relative differential number densities of nonthermal $^{16}$O ions at various downstream positions in an SNR at the beginning of the radiative stage, for $E_{\rm SN}=1.5\times 10^{51}$~erg, $n_{\rm H}=10$~cm$^{-3}$, and $s_p=4.3$. The dashed lines show the results obtained at $R=0.97R_s$ and $0.7R_s$ without taking into account the Coulomb losses.}
\label{fig1}
\end{figure}

The main energy loss processes that can affect the energy distribution of low-energy hadronic CRs during their transport in the downstream medium are adiabatic and Coulomb cooling. The formalism outlined in Appendix~A allows us to calculate the initial energy $E_i$ at time $t_i$ of the particle $j$ that has an energy $E$ at time $t$ and position $R$ (from eqs.~\ref{eqa6} and \ref{eqa8} for $E < E_t=200$~MeV~nucleon$^{-1}$ and eq.~\ref{eqa13} for $E \geq E_t$). Appendix~A also gives the transfer function $G$ such that
\begin{equation} 
n_{j} \left(E,R,t\right)=n_{j} \left(E_{i} ,R_{s} (t_{i} ),t_{i} \right) G\{E,E_i,\rho_{\rm gas}[R(t)]/\rho_{\rm CSM}\}~.    
\label{eq14} 
\end{equation} 
Remarkably, the $G$ function depends on the relative density of the downstream thermal gas, $\rho_{\rm gas} [R(t)]/\rho_{\rm CSM}$, obtained from the self-similar Sedov solution, but not explicitly on the nature of the particle $j$ (see eqs.~\ref{eqa11} and \ref{eqa15}). However, the relation between $E_i$ and $E$ that takes into account the Coulomb losses depends on the atomic and mass numbers of the fast ions (eq.~\ref{eqa6}). 

Figure~\ref{fig1} shows calculated differential number densities of fast $^{16}$O ions at various radii in an SNR entering the radiative phase. The effect of the Coulomb losses is apparent below $\sim 100$~MeV~nucleon$^{-1}$. The effect of the adiabatic losses can be assessed by comparing the differential densities above $200$~MeV~nucleon$^{-1}$. Noteworthy, these losses are partly offset by the fact that the CR injection rate decreases with time (eq.~\ref{eq12}). Thus, the layer being at the position $R=0.7R_s$ at time $t_{\rm rad}$ was shocked at $t_i=646$~yr, and the differential CR injection rate at that early time was much higher: $Q_s(E,t_i)=14.6\times Q_s(E,t_{\rm rad})$. In comparison, at $t=t_{\rm rad}$, the density of fast $^{16}$O ions of energy $E>200$~MeV~nucleon$^{-1}$ is about 1.6 times lower in this layer than that at the forward shock position (Figure~\ref{fig1}). 

\begin{figure}
\centering
\includegraphics[width=7.2cm]{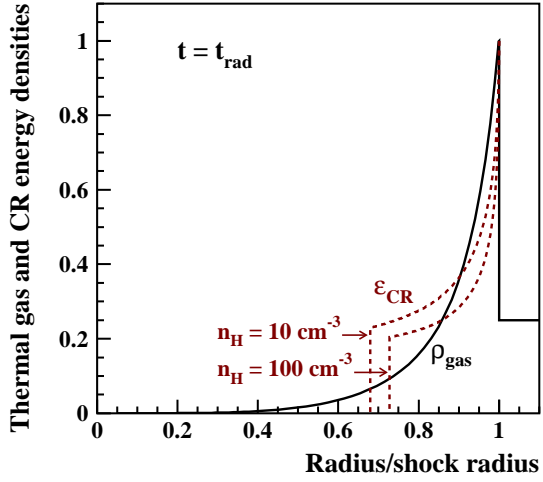}
\caption{Radial profiles of the thermal gas density ($\rho_{\rm gas}$) and the energy density in nonthermal protons ($\epsilon_{\rm CR}$) in an SNR at the end of the Sedov-Taylor phase. $\rho_{\rm gas}$ is the self-similar Sedov solution; $\epsilon_{\rm CR}$ is calculated with $E_{\rm SN}=1.5\times 10^{51}$~erg, $M_{\rm ej}=10~M_\odot$, $s_p=4.3$, and for two values of $n_{\rm H}$.}
\label{fig2}
\end{figure}

In Figure~\ref{fig2} we show radial profiles of CR proton energy densities obtained by integration of $n_p(E,R,t_{\rm rad})$ over energy. The faster decay of $\epsilon_{\rm CR}(R,t_{\rm rad})$ for $n_{\rm H}=100$~cm$^{-3}$ than for $n_{\rm H}=10$~cm$^{-3}$ shows the significance of Coulomb energy losses for such densities of the CSM. The sudden drop of $\epsilon_{\rm CR}(R,t_{\rm rad})$ at $R/R_s=0.68$ (for $n_{\rm H}=10$~cm$^{-3}$) and $R/R_s=0.728$ (for $n_{\rm H}=100$~cm$^{-3}$) is due to the assumption of the beginning of the CR acceleration at the time $t_{\rm ST}$ marking the start of the Sedov-Taylor phase (eq.~\ref{eq2}): this limit is the maximum distance traveled by CRs in the downstream medium during the time interval $t_{\rm rad} - t_{\rm ST}$. We see in this Figure than most of the thermal gas and CRs are located close to the blast wave, which shows the limits of previous models that have considered homogeneous SNRs \citep[e.g.][]{par99a,par99b}. 

\subsection{Light element production}

The production rate of a light element $k$ by CR interaction with the ambient gas at radius $R$ and time $t$ (in atoms~cm$^{-3}$~s$^{-1}$) is given by
\begin{equation} 
q_k \left(R,t\right)=\sum _{i,j}\int _{0}^{\infty} n_i(R,t) n_j(E,R,t) \sigma _{i,j;k}(E) v_j(E)dE~,  
\label{eq15} 
\end{equation} 
where $v_j(E)$ is the velocity of the CR particle of type $j$, $\sigma _{i,j;k} (E)$ the cross section for the nuclear reaction $i+j \rightarrow k$, and $n_i(R,t)=n_{\rm H} x_i \rho_{\rm gas}[R(t)]/\rho_{\rm CSM}$, with $x_i$ the abundance of the constituent $i$ in the CSM, which we assume to have the solar system composition \citep{asp09}. We use for the accelerated particles the composition of the current epoch Galactic CRs at their sources, which we obtain by taking the heavy ion abundances relative to O from \citet{eng90} and the alpha-to-oxygen and proton-to-alpha abundance ratios recommended by \citet{mey97}: $C_\alpha/C_{\rm O}=19$ and $C_p/C_\alpha=15$. The resulting CR source composition is consistent with the theoretical works of \citet{put11}. We assume that all the accelerated ion species have the same energy spectrum as a function of kinetic energy per nucleon. 

In the next section we present with some details the synthesis of $^{10}$Be by CR interaction in an SNR during the Sedov-Taylor phase. We then evaluate in Sect.~2.3.2 the significance of the production of the stable light elements $^{6,7}$Li, $^9$Be, and $^{10,11}$B by this process.

\subsubsection{$^{10}{\rm Be}/^9{\rm Be}$ at the end of the Sedov phase}

\begin{figure}
\centering
\includegraphics[width=7.2cm]{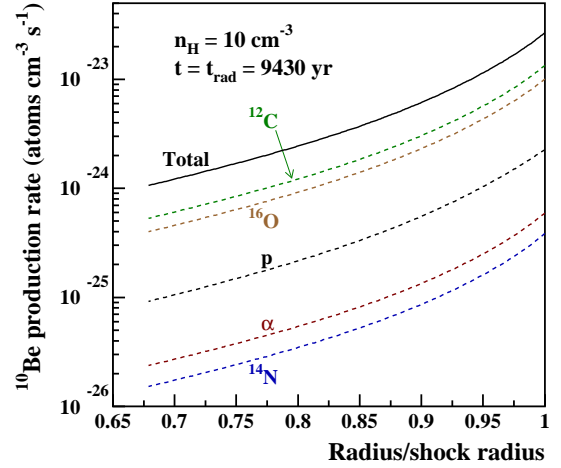}
\caption{$^{10}$Be production rates at the end of the Sedov-Taylor phase as a function of the downstream position $R/R_s$, for $E_{\rm SN}=1.5\times 10^{51}$~erg, $n_{\rm H}=10$~cm$^{-3}$, and $s_p=4.3$. The contributions of fast protons, $\alpha$ particles, $^{12}$C, $^{14}$N, and $^{16}$O ions to the total rate are shown by the dashed curves.}
\label{fig3}
\end{figure}

\begin{figure*}
\centering
\includegraphics[width=0.8\linewidth]{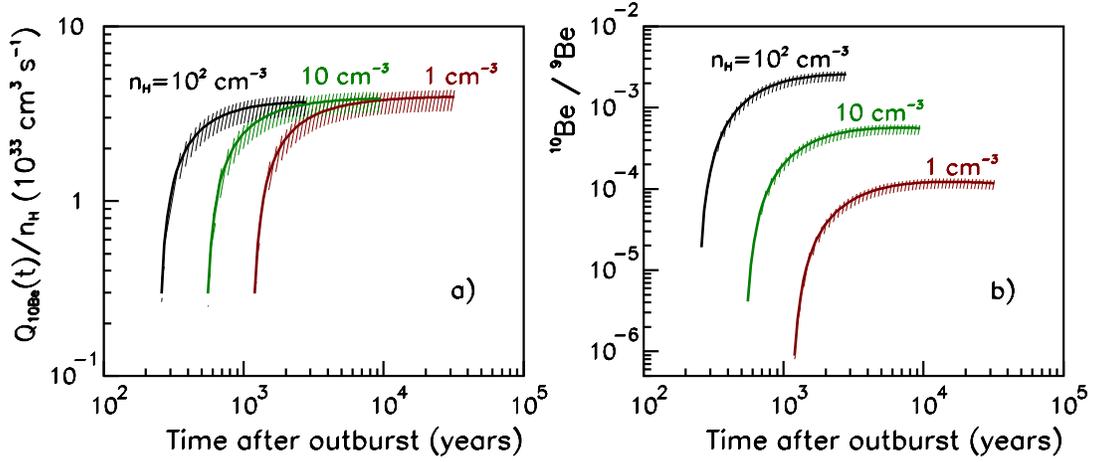}
\caption{Temporal evolution of the production of $^{10}$Be in an SNR during the Sedov-Taylor phase. \textit {Left}: instantaneous production rate (in atoms s$ ^{-1}$) divided by the H density in the CSM ($n_{\rm H}$); \textit{right}: evolution of the abundance ratio $^{10}{\rm Be} / ^9{\rm Be}$. The calculations were carried out with $E_{\rm SN}=1.5\times 10^{51}$~erg and $M_{\rm ej}=10~M_\odot$ (see eq.~\ref{eq2}), and three different values of $n_{\rm H}$. The hatched areas reflect the errors arising from the uncertainty in the slope of the phase-space distribution: $s_p=4.3 \pm 0.1$ (eq.~\ref{eq7}).}
\label{fig4}
\end{figure*}

$^{10}$Be is mostly produced by spallation of CNO nuclei in collisions with protons and $\alpha$ particles. The corresponding reaction cross sections are presented in Appendix~B. In Figure~\ref{fig3}, we show total production rate of $^{10}$Be atoms calculated from eq.~(\ref{eq15}), together with the contributions of the most abundant CR species to this production . We see that $^{10}$Be is mainly synthesized by spallation of fast $^{12}$C and $^{16}$O, whose abundances in the CR source composition are $C_{\rm ^{12}C}/C_p=2.8\times10^{-3}$ and $C_{\rm ^{16}O}/C_p=3.5\times10^{-3}$, respectively. In comparison, the abundances of these species in the solar system composition are $x_{\rm ^{12}C} = 2.92\times10^{-4}$ and $x_{\rm ^{16}O} = 5.36\times10^{-4}$. 

$^{10}$Be nuclei produced by spallation of fast heavy ions acquire in the reactions a recoil kinetic energy per nucleon similar to that of the projectiles, i.e. $E_{\rm rec} \gsim 30$~MeV~nucleon$^{-1}$ and $\gsim 10$~MeV~nucleon$^{-1}$ for spallation in H and He, respectively (see Figure~\ref{figb1} in Appendix~B). Depending on the plasma density in the SNR, these fast $^{10}$Be ions may not get thermalized during the Sedov-Taylor stage and some of them may escape into the ISM during the radiative phase. To estimate the importance of this loss, we assume that the escaping $^{10}$Be ions are those having an energy loss time at the time of their production, $\tau_{\rm loss} \sim E_{\rm rec} / [(dE/dt)_{\rm Coul}(E_{\rm rec})]$, greater than the duration of the Sedov phase, $t_{\rm rad}-t_{\rm ST} \approx t_{\rm rad}$ (eq.~\ref{eq4}). Using the power-law approximation of the Coulomb energy loss rate presented in Appendix~A (eq.~\ref{eqa1}), we readily obtain a lower limit on the recoil energy of the escaping $^{10}$Be: 
\begin{equation} 
E_{\rm rec} > \left( K X_{\rm comp} n_{\rm H} Z^2 t_{\rm rad} / A \right)^{1 \over 1+\beta}~,
\label{eq16} 
\end{equation} 
where $\beta=0.47$ and $K=7.5\times10^{-12}$~MeV~nucleon$^{-1}$~s$^{-1}$~cm$^3$ (see Appendix~A), $Z=4$, $A=10$, and $X_{\rm comp} \sim 2$, the latter factor taking into account the higher density of the SNR shell relative to that of the CSM due to the shock compression ($n_{\rm H} \equiv n_{\rm H}^{\rm CSM}$ in the above equation). With these numerical values and equation~(\ref{eq4}), we get:
\begin{equation} 
E_{\rm rec} > \left(8.2 {\rm \; MeV~nucleon^{-1}}\right)\left(\frac{E_{\rm SN} }{10^{51} {\rm \; erg}} \right)^{0.16} \left(\frac{n_{\rm H} }{1{\rm \; cm}^{{\rm -3}} } \right)^{0.32}~.
\label{eq17} 
\end{equation} 
This result shows that most $^{10}$Be nuclei produced in inverse kinematics, i.e. by spallation of fast heavy ions, will escape the SNR during the radiative phase. On the other hand, as illustrated in Figure~\ref{figb2}, the vast majority of $^{10}$Be ions synthesized by accelerated proton- and $\alpha$-particle-induced spallations (direct kinematics) are produced with recoil energy $E_{\rm rec}< 8$~MeV~nucleon$^{-1}$, and will stop in the SNR in a time $\tau_{\rm loss} < t_{\rm rad}$ as long as $n_{\rm H} \gsim 1$~cm$^{-3}$. Here we make the simplifying assumption that all $^{10}$Be produced in direct kinematics are trapped in the SNR and all those produced in inverse kinematics are lost in the ISM before the end of the SNR. Because the latter do not enrich the shocked gas layer, they are no longer considered in this work. However, we see in Figure~\ref{fig3} that lost $^{10}$Be ions are about 10 times more abundant than those produced in direct kinematics (mainly by proton reactions). 

$^{10}$Be ions produced in a downstream region near the blast wave may be able to cross the shock front to the upstream medium and thus be injected in the diffusive shock acceleration process. This may occur in a region of thickness 
\begin{equation} 
\Delta R \sim {r D \over V_s}~,
\label{eq18} 
\end{equation} 
where $r$ is the shock compression ratio ($r=4$ for a test-particle strong shock) and $D$ is the spatial diffusion coefficient of $^{10}$Be nuclei in the downstream medium. Taking for the latter the Bohm value, $D = v r_g /3$ where $v$ is the particle speed and $r_g=pc/QeB_{\rm ps}$ the particle gyroradius ($p$ is the particle momentum, $Q$ the charge number, $-e$ the electronic charge, and $B_{\rm ps}$ the postshock magnetic field), we get for fully stripped ($Q=4$), non-relativistic $^{10}$Be ions:
\begin{eqnarray}
\frac{\Delta R}{R_{s} } &=& 2.8\times 10^{-8} \left(\frac{E_{{\rm rec}} }{8{\rm \; MeV\; nucleon}^{{\rm -1}} } \right)\left(\frac{B_{\rm ps} }{100{\rm \; }\mu {\rm G}} \right)^{-1} \times \nonumber \\
& &
\left(\frac{E_{\rm SN} }{10^{51} {\rm \; erg}} \right)^{-2/5} \left(\frac{n_{\rm H} }{1{\rm \; cm}^{{\rm -3}} } \right)^{2/5} \left(\frac{t}{10^{4} {\rm \; yr}} \right)^{1/5}~.  
\label{eq19} 
\end{eqnarray} 
Given the small size of the region of interest, the reacceleration of $^{10}$Be ions produced by spallation of ambient CNO by fast protons and $\alpha$ particles can be safely neglected. 

Figure~\ref{fig4}a shows the quantity
\begin{equation} 
{Q_{^{10}{\rm Be}} (t) \over n_{\rm H}} ={1 \over n_{{\rm H}}} \int _{0}^{R_s(t)} q_{^{10}{\rm Be}}(R,t) \times 4 \pi R^2 dR 
\label{eq20} 
\end{equation} 
for three values of $n_{\rm H}$. Here, $q_{^{10}{\rm Be}}(R,t)$ is calculated from equation~(\ref{eq15}) with $j \equiv p$ or $\alpha$. We see that the $^{10}$Be production rate rapidly increases at the beginning of the Sedov-Taylor phase and then becomes nearly independent of time. This is because the contribution of freshly accelerated CRs to the total $^{10}$Be production becomes negligible at the end of this stage. At $t_{\rm rad}$,  $Q_{^{10}{\rm Be}} / n_{\rm H}$ is in the range $(3$~--~$4)\times 10^{33}$~cm$^3$~s$^{-1}$ whatever the CSM density. 

Figure~\ref{fig4}b shows the temporal evolution of the isotopic ratio $^{10}{\rm Be} / ^9{\rm Be}$ in the SNR interior: 
\begin{equation} 
\frac{{}^{10} {\rm Be}}{{}^{9} {\rm Be}} (t)=\frac{\int _{0}^{t}Q_{{}^{{\rm 10}} {\rm Be}} \left(t'\right)dt' }{x_{{}^{9} {\rm Be}} V_{\rm SNR} (t)n_{\rm H} }~,     
\label{eq21} 
\end{equation} 
where $V_{\rm SNR}(t)=4\pi R_s^3(t)/3$ is the SNR volume at time $t$ and $x_{^9{\rm Be}}=2.63\times 10^{-11}$ is the estimated protosolar abundance of $^9$Be \citep{asp09}. This equation assumes that the number of $^9$Be atoms synthesized in the SNR can be neglected in front of the number of circumstellar atoms collected by the forward shock. This will be checked in the next section. 

The mass of material contained in the SNR at the end of the Sedov-Taylor phase is significant: $M_{\rm SNR}=\rho_{\rm CSM}V_{\rm SNR}(t_{\rm rad})=1460$, $840$, and $490~M_\odot$, for $n_{\rm H}=1$, $10$, and $100$~cm$^{-3}$, respectively. At that time, the isotopic ratio $^{10}{\rm Be} / ^9{\rm Be}$ in this material amounts to respectively $\sim 1.2\times 10^{-4}$, $\sim 5.6\times 10^{-4}$, and $\sim 2.5\times 10^{-3}$ (Figure~\ref{fig4}b). In comparison, the inferred $^{10}{\rm Be} / ^9{\rm Be}$ ratio in the protosolar molecular cloud is  $\sim 3\times 10^{-4}$ \citep{wie12}. The implications of these calculations for the astrophysical context of solar system formation are discussed in Section~3. 

\subsubsection{Production of stable light elements}

We have calculated the production of the stable isotopes $^6$Li, $^7$Li, $^9$Be, $^{10}$B, and $^{11}$B, from the reaction cross sections discussed in \citet{tat07} and using the same formalism as for $^{10}$Be. In particular, since our first goal is to study the composition of the protosolar molecular cloud, we have considered only the reactions induced by accelerated protons and $\alpha$ particles. Some results obtained with $n_{\rm H}=100$~cm$^{-3}$ are given in Table~\ref{tab1}. 

\begin{deluxetable}{lcc}
\tablewidth{0.75\linewidth}
\tablecaption{Production of stable light elements in an SNR\tablenotemark{a}\label{tab1}}
\tablehead{
\colhead{Isotope}   &  \colhead{${Q_k(t_{\rm rad}) / n_{\rm H}}$\tablenotemark{b}} & \colhead{$N_k/(x_kn_{\rm H}V_{\rm SNR})$\tablenotemark{c}} \\
\colhead{}          &  \colhead{($10^{34}$~atoms~cm$^3$~s$^{-1}$)} & \colhead{(\textperthousand)}}
\startdata
$^6$Li    & $10.0$ &  $15.8$\tablenotemark{d} \\
$^7$Li    & $17.7$ &  $2.4$\tablenotemark{d} \\
$^9$Be    & $1.1$  &  $7.5$ \\
$^{10}$B  & $4.3$  &  $6.6$ \\
$^{11}$B  & $11.0$ &  $4.3$
\enddata
\tablenotetext{a}{For $n_{\rm H}=100$~cm$^{-3}$, $s_p=4.3$, $E_{\rm SN}=1.5\times 10^{51}$~erg, and $M_{\rm ej}=10~M_\odot$.}
\tablenotetext{b}{Instantaneous production rate of the isotope $k$ at $t_{\rm rad}$ divided by the H density in the CSM. This ratio depends only weakly on $n_{\rm H}$ (see Figure~\ref{fig4}a for $^{10}$Be).}
\tablenotetext{c}{Number of atoms $k$ synthesized in the SNR during the Sedov phase ($N_k$) divided by the number of atoms $k$ of the CSM collected by the forward shock during the same period, in per mil.}
\tablenotetext{d}{Escape from the SNR of $^6$Li and $^7$Li produced in the $\alpha + \alpha$ reaction is not taken into account (see text).}
\end{deluxetable}

$^6$Li and $^7$Li are mainly synthesized in collisions of accelerated $\alpha$ particles with ambient He. In this so-called $\alpha + \alpha$ reaction, a significant fraction of Li nuclei are produced with recoil energies $> 8$~MeV~nucleon$^{-1}$ and will not get thermalized in the SNR (see eq.~\ref{eq16}). The relative numbers of $^6$Li and $^7$Li given in Table~\ref{tab1} thus represent upper limits to the possible contamination of the protosolar molecular cloud by Li atoms synthesized in an SNR. Anyway, we see that the production of stable elements by this process is negligible compared to the composition of the protosolar molecular cloud resulting from billions of years of Galactic chemical enrichment. This confirms previous results which showed that nonthermal production of the stable isotopes of Li, Be and B in SNRs is not important for Galactic chemical evolution \cite[e.g.,][]{par99a,par99b,tat11}. 

\section{The origin of $^{10}$Be in FUN-CAIs}

The model developed in Section~2 shows that production of $^{10}$Be in an SNR can be significant when compared to the level of radioactive contamination of the early solar system. This result prompted us to study in detail the origin of ${\rm ^{10}Be}$ in FUN-CAIs. In Section~3.1, we first examine if the ${\rm ^{10}Be}$ nuclei incorporated in these objects were produced within the solar system by spallogenic nucleosynthesis induced by solar energetic particles, or if they were inherited from the protosolar molecular cloud, as suggested by \citet{wie12}. Reaching the conclusion that the ${\rm ^{10}Be}$ found in FUN-CAIs was not produced within the early solar system, we then study in Section~3.2 the origin of this radioactivity in the protosolar molecular cloud. In Sections~3.2.1 to 3.2.5, we successively review five models of $^{10}$Be enrichment of a molecular cloud that could potentially account for the ${\rm ^{10}Be} / {\rm ^9Be}$ ratio recorded by FUN-CAIs. 

\subsection{$^{10}$Be in FUN-CAIs: in situ production within the solar system or a protosolar heritage?}

The variability of $^{10}$Be abundances measured in canonical CAIs and refractory hibonites supports an heterogeneous distribution of this radioactivity in the early solar system, which is best explained by a local production by solar energetic particles \citep[][and references therein]{sri13}. In this context, $^{10}$Be could have been synthesized in three types of setting: (i) through spallation of solar nebula gas close to the proto-sun \citep[e.g.][]{mck00}, (ii) by large flares in the atmosphere of the young star \citep{bri10}, or (iii) by irradiation of the refractory inclusions themselves or their precursor materials in the protoplanetary disk \citep{gou01,gou13,liu09,liu10}. 

The initial abundances of $^{10}$Be measured by \citet{wie12} in two FUN-CAIs are  ${\rm ^{10}Be} / {\rm ^9Be} = (2.77\pm0.24)\times 10^{-4}$ in AXCAI 2771 (a CAI from the CV chondrite Axtell) and ${\rm ^{10}Be} / {\rm ^9Be} = (3.37\pm0.20)\times 10^{-4}$ in KT1 (a CAI from the CV chondrite NWA 779). These abundances are lower than those found in canonical CAIs and hibonites \citep[see][]{sri13}. FUN-CAIs are known to contain large mass-independent isotope anomalies of nucleosynthetic origin \citep{bir04}, which reflects that these objects had little isotope exchange with a gas of solar composition. Thus, as already pointed out by \citet{wie12}, $^{10}$Be formation in any solar gas cannot explained the presence of $^{10}$Be in FUN-CAIs.  We argue below that an in situ production of $^{10}$Be by irradiation of the FUN CAIs themselves is also unlikely.

\subsubsection{$^6$Li overproduction in irradiated FUN-CAIs}

\citet{wie12} suggested that an in situ spallogenesis of $^{10}$Be should be accompanied by a significant production of $^6$Li. To provide a quantitative discussion of this argument, we first assume that the differential flux spectrum of the CAI-irradiating solar energetic particles is a power law in kinetic energy per nucleon above the energy thresholds of the $^6$Li- and $^{10}$Be-producing nuclear reactions. We allow the spectral index of this power-law distribution to vary between $s=2.5$ and $4$, which is typical of modern solar flares \citep[see also][]{dup07}. We take the accelerated $\alpha$-particle to proton abundance ratio to be $\alpha/p=0.1$. Considering that in refractory solids $^6$Li and $^{10}$Be are mainly produced by spallation of target $^{16}$O, we obtain that the production rate ratio $Q({\rm ^6Li})/Q({\rm ^{10}Be})$ ranges from 26 for $s=2.5$ to 62 for $s=4$. 

\citet{wie12} carried out most of their analyses on melilite samples. They measured Be concentrations in melilite grains of the FUN-CAIs ranging from 54 to 1214~ppb, with an mean value of ${\rm [Be]}=376$~ppb. The mean concentration of spallogenic $^6$Li produced in situ is then predicted to be
\begin{equation} 
[^6{\rm Li}]_{\rm spal} ={Q({\rm ^6Li}) \over Q({\rm ^{10}Be})} \times {{\rm ^{10}Be} \over {\rm ^9Be}} \times {\rm [Be]} \in [2.9~{\rm ppb},7.0~{\rm ppb}]~,
\label{eq22} 
\end{equation} 
where ${\rm ^{10}Be} / {\rm ^9Be} \sim 3\times 10^{-4}$ is the approximate initial abundance of $^{10}$Be  in the FUN-CAIs \citep{wie12}. The mean concentration of spallogenic $^7$Li can be calculated in the same way. The resulting Li isotopic ratio ranges from 0.86 for $s=4$ to 1.24 for $s=2.5$. 

The measured mean concentration of $^6$Li in melilite of the FUN-CAIs if 10.9~ppb \citep{wie12}. Assuming that it results from a mixing of spallogenic and chondritic Li (with the chondritic ratio $^7{\rm Li} / ^6{\rm Li}=12.06\pm 0.03$; \citealp{sei07}), we would expect a mean isotopic ratio in the samples ranging from $^7{\rm Li} / ^6{\rm Li}=4.9$ for $s=4$ to $^7{\rm Li} / ^6{\rm Li}=9.2$ for $s=2.5$. In comparison, the weighted mean of the Li measurements in melilite of the FUN-CAIs is $^7{\rm Li} / ^6{\rm Li}=11.93 \pm 0.02$ \citep{wie12}, much closer to the chondritic value. 

The issue of spallogenic $^6$Li overproduction is even more significant when considering individual samples with high Be abundance and/or low $^6$Li concentration. For example, \citet{wie12} measured a Be concentration of $1214 \pm 122$~ppb in one melilite grain of the FUN-CAI KT1. In the model of in situ production, this sample should contain a concentration of spallogenic $^6$Li in the range 9.5~--~22.6~ppb (eq.~\ref{eq22}), whereas the measured $^6$Li concentration is much lower: $[^6{\rm Li}]=2.19\pm 0.25$~ppb. As \citet{wie12} already noted \citep[see also][]{thr08}, it is unlikely that the original Li content of this FUN-CAI was completely erased during secondary events and replaced by Li from a chondritic reservoir. 

\subsubsection{Inconsistent in-situ production of $^{10}{\rm Be} / ^{9}{\rm Be}$ in different mineral phases}

In fact, regardless of the Li data, that samples with very different Be concentrations fall into the same isochron in a $^{10}$Be--$^{10}$B isochron diagram strongly argues against an in-situ production of $^{10}$Be. In the model of spallogenesis in refractory solids, the initial abundance of $^{10}$Be in a given CAI is generally  calculated as \citep[see, e.g.,][]{gou13}
\begin{equation} 
{{\rm ^{10}Be} \over {\rm ^9Be}} = {[^{16}{\rm O}] \over [{\rm Be}]} \int _{0}^{\infty} N_p(E) \left[\sigma _p(E) + {\alpha \over p}\sigma _\alpha(E)\right]dE~.  
\label{eq23} 
\end{equation} 
Here, $\sigma _i$ is the cross section for the reaction $i + ^{16}{\rm O} \rightarrow ^{10}{\rm Be}$ and $N_p(E)$ is the differential fluence of energetic protons (number of particles cm$^{-2}$ MeV$^{-1}$). The concentration of $^{16}$O in CAI minerals is about 60\% (by number of atoms). But [Be] varies considerably from one mineral phase to another. Thus, the mean Be concentration in the pyroxene grains of the FUN-CAI KT1 is $[{\rm Be}]=38$~ppb, which is an order of magnitude lower than in melilite \citep{wie12}. The fact that the initial abundance of $^{10}$Be was about the same in these two mineral phases --~as can be deduced from the  $^{10}$Be--$^{10}$B isochron diagram \citep[][Figure~1b]{wie12}~-- would require the proton fluence received by melilite to be about ten times higher than that received by pyroxene (see eq.~\ref{eq23}), which is truly unlikely. In other words, for a given proton fluence the $^{10}{\rm Be} / ^{9}{\rm Be}$ ratio should be ten times higher in pyroxene than in melilite, which is not observed. 
 
Noteworthy, this argument is also valid for canonical CAIs, which makes a case against in situ spallogenesis also for these objects. But in contrast to FUN-CAIs, canonical CAIs may have inherited $^{10}$Be from an irradiated solar nebula at the inner edge of the accretion disk. In \citet{sri13}, the nebula is considered to be a mixture of a small fraction of refractory solids and of un-fractionated solar gas, with a solid-to-gas mass ratio of $\sim 10^{-3}$--$10^{-2}$.  

\subsubsection{Overheating of irradiated FUN-CAIs}

Another argument against an in situ spallogenic origin of $^{10}$Be in FUN-CAIs is related to the heating of solids irradiated by high particle fluxes. Applying equation~(\ref{eq23}) with ${\rm ^{10}Be} / {\rm ^9Be} = 3\times 10^{-4}$ and ${\rm [Be]}=376$~ppb (the mean concentration measured by \citet{wie12} in melilite), we obtain a required  fluence in protons of energy greater than 10~MeV ranging from $F_p(>10~{\rm MeV})=1.2\times 10^{18}$~cm$^{-2}$ for $s=2.5$ to $5.4\times 10^{18}$~cm$^{-2}$ for $s=4$.  At energies below the thresholds of the nuclear reactions, we assume that the accelerated particle spectrum extends as a power-law of index $s$ between 2.5 and 4 down to 1~MeV~nucleon$^{-1}$ and is flat below ($s=0$). Such a spectral shape is approximately representative of that measured in modern solar flares \citep{rea99}. Taking into account accelerated $\alpha$ particles with the same source spectrum as for protons and with $\alpha/p=0.1$, the required total energy fluence in fast particles becomes $F_{\rm tot}=3.2\times10^{14}$~erg~cm$^{-2}$ for $s=2.5$ and $F_{\rm tot}=3.6\times10^{16}$~erg~cm$^{-2}$ for $s=4$. 

To estimate what fraction of this nonthermal energy was deposited in the FUN-CAIs (in the scenarion of in-situ spallogenesis), we first evaluated the projected range of protons and $\alpha$ particles in these solids using the SRIM code (Stopping and Range of Ions in Matter; \citealp{zie10}). We found that fast particles of energy less than 21~MeV~nucleon$^{-1}$ should stop in 2.5~mm of this material, which is the approximate diameter of the FUN-CAIs KT1 and AXCAI 2771 \citep{thr08,sri00}. The energy fluence deposited in the solids by particles of energy $E<21$~MeV~nucleon$^{-1}$ is then found to be $F_{\rm dep}=2.6\times10^{14}$~erg~cm$^{-2}$ for $s=2.5$ and $F_{\rm dep}=3.6\times10^{16}$~erg~cm$^{-2}$ for $s=4$ (for the steepest particle spectrum, 99.9\% of the total nonthermal energy is contained in particles of less than 21~MeV~nucleon$^{-1}$). 

The heating of the target depends on the instantaneous power they receive and thus on the irradiation time. The typical duration of X-ray flares of pre-main-sequence stars ranges from a few hours to a few days, with a mean value of $\Delta t \approx 10^5$~s \citep{wol05}. Assuming that the required energy fluence is delivered by one of such flares would give a mean energy flux endured by the refractory targets ranging from $f_{\rm dep}=2.6\times10^9$ to $3.6\times10^{11}$~erg~cm$^{-2}$~s$^{-1}$, which corresponds to an equilibrium temperature in the range $T=2600$--$8930$~K ($f_{\rm dep} = \sigma T^{4}$ where $\sigma =5.67\times10^{-5}$~erg~cm$^{-2}$~s$^{-1}$~K$^{-4}$ is the Stephan-Boltzman constant). Under such a power deposition, the targets would rapidly evaporate. 

FUN-CAIs contain large nucleosynthetic anomalies in elements such as barium and strontium, whose melting temperatures are 1000~K and 1050~K, respectively. It proves that these objects were not heated by secondary processes above $\sim 1000$~K and therefore did not receive at any time a particle energy flux larger than $f_{\rm dep}=5.7\times10^7$~erg~cm$^{-2}$~s$^{-1}$. In the model of in-situ $^{10}$Be production, targets were then necessarily exposed to a large number of relatively weak flares, $N>F_{\rm dep} / (\Delta t f_{\rm dep})=46$ for $s=2.5$ ($N>6316$ for $s=4$). This result is difficult to reconcile with the popular X-wind model that assumes an intense irradiation of bare solids in the reconnection ring close to the proto-sun \citep[see][]{gou01}. 

Thus, in view of these issues, in situ spallogenesis seems unable to account for the presence of $^{10}$Be in FUN-CAIs. As these objects had minor interaction with any solar gas, the $^{10}{\rm Be} / ^{9}{\rm Be}$ ratio that they recorded must reflect the level of $^{10}$Be contamination of the protosolar molecular cloud. 

\subsection{The origin of $^{10}$Be in the protosolar molecular cloud}

\begin{figure}
\centering
\includegraphics[width=8.3cm]{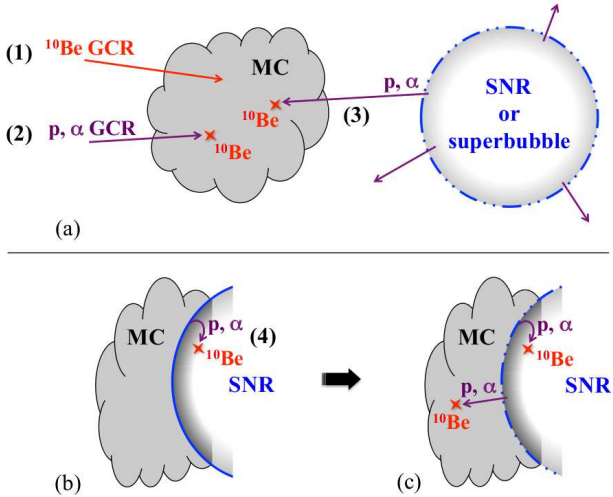}
\caption{Schematic illustration of various models of $^{10}$Be enrichment of an isolated molecular cloud ({\it upper panel}, {\it a}) and of a molecular cloud impacted by an SNR ({\it lower panels}, {\it b} and {\it c}): ($1$) trapping of $^{10}$Be GCRs, ($2$) spallation reactions induced by GCR protons and $\alpha$-particles, ($3$) spallation reactions induced by protons and $\alpha$-particles escaped from an isolated SNR in the radiative phase or from a superbubble formed by the activity of several massive stars and SNe in an OB association, and ($4$) spallation reactions induced by CRs trapped within a SNR in the Sedov-Taylor phase. In panel ({\it c}) the SNR has reached the radiative stage and $^{10}$Be nuclei are produced by freshly accelerated CRs diffusing both upstream and downstream the shock. }
\label{schema10be}
\end{figure}

Several SNRs have been detected in gamma-rays at both TeV and GeV energies \cite[see][and references therein]{hel12}. In most of these objects, the gamma-ray emission is thought to be produced by interaction of trapped CRs with shocked material inside the remnant. In two cases, however, part of the gamma-ray emission is clearly coming from outside the remnant and is most likely due to CRs that escaped the acceleration site and now interact with nearby molecular clouds: W28 \citep{abd10} and W44 \citep{uch12}. Both SNRs are $10^4$--$10^5$~yr old and have probably reached the radiative stage of their evolution. 

Inspired by these gamma-ray observations, we study in the following various scenarios that could potentially explain the $^{10}$Be abundance of the protosolar molecular cloud. In the first one (Sect.~3.2.3), we consider a SN exploding in the intercloud medium and releasing non-relativistic CRs at the end of the Sedov-Taylor phase. The escaping particles then spread into the surrounding ISM and produce significant $^{10}$Be nuclei in a nearby molecular cloud. In the continuation of this model, we consider in Section~3.2.4 the irradiation of a giant molecular cloud by CRs produced by an ensemble of SNe exploding within a superbubble formed by an OB association. In Section~3.2.5, we then consider a scenario where a massive star escapes from its parent OB association and explodes within a molecular cloud complex, so that some molecular gas is impacted by the remnant during the Sedov phase. Some $^{10}$Be is then produced by trapped CRs interacting with shocked material. But before that, we discuss two other models for the origin of protosolar $^{10}$Be already studied by \citet{des04}: trapping of GCRs in the collapsing protosolar cloud core (Sect.~3.2.1) and irradiation of the presolar molecular cloud by background GCRs (Sect.~3.2.2). The five models considered in the following are shown schematically in Figure~\ref{schema10be}.

\subsubsection{Trapping of $^{10}$Be GCRs}

\begin{figure}
\centering
\includegraphics[width=7.0cm]{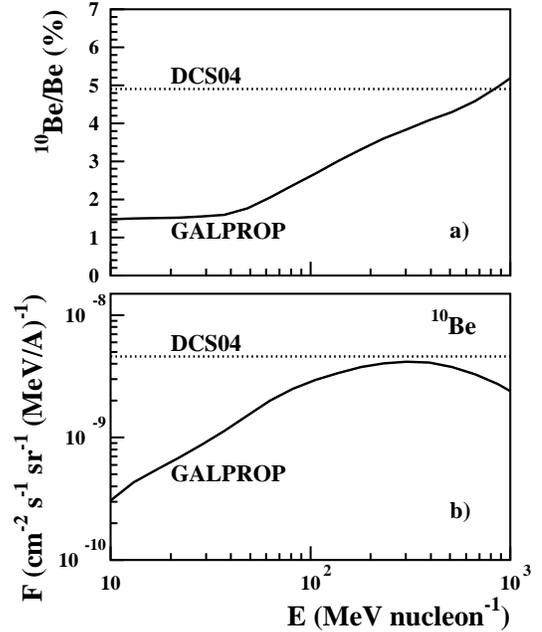}
\caption{$^{10}{\rm Be} / {\rm Be}$ ratio ({\it a}) and $^{10}{\rm Be}$ flux ({\it b}) in current-epoch GCRs. Here, ${\rm Be} \equiv ^7$Be+$^{9}$Be+$^{10}$Be. The GALPROP model is a conventional diffusive reacceleration model \citep{str01,ptu06} providing a good description of elemental energy spectra measured with the {\it Advanced Composition Explorer} \citep{lav13}. The dotted lines show the values adopted by \citet{des04}.}
\label{fig5}
\end{figure}

Based on the work of \citet{des04}, \citet{wie12} suggested that the $^{10}$Be abundance in the protosolar molecular cloud was generated by enhanced trapping of GCRs. This model was first proposed to explain the presence of $^{26}$Al in meteoritic inclusions \citep{cla95}. Although the hypothesis of GCR trapping fall short to explain the high $^{26}$Al content of the early solar system \citep{lee98}, it is worth considering in the case of $^{10}$Be in view of the high abundance of this secondary nucleus in the GCRs. 

As a molecular cloud core collapses, its column density rises and finally reaches values in the order of the stopping length of energetic ions ($\sim 0.01$~g~cm$^{-2}$ for ions a few tens of MeV per nucleon). \citet{des04} studied this mechanism within the framework of the model of \citet{des01} for the evolution of the molecular cloud core magnetic field and density. They consider a timescale for the entire process of a few Myr, an absolute flux of GCRs at the time of the solar system formation two times higher than the present one, and a $^{10}$Be/H ratio in the GCRs slightly lower (by 83\%) than the present value. Under these hypotheses and with a present-day GCR flux of $^{10}$Be nuclei at energies $\lsim 100$~MeV~nucleon$^{-1}$ of $4.6 \times 10^{-9}$~cm$^{-2}$~s$^{-1}$~sr$^{-1}$~(MeV/nucleon)$^{-1 }$, \citet{des04} concluded that a $^{10}{\rm Be}/^{9}{\rm Be}$ ratio as high as $\sim 10^{-3}$ can be obtained by trapping of energetic $^{10}$Be.

We have revisited the trapping mechanism using the GCR propagation code GALPROP\footnote{\url{http://galprop.stanford.edu/}} to better estimate the energy dependence of the $^{10}$Be flux at low energies. We adopted a conventional diffusive reaccelerating model (\citealp{str01,ptu06}; parameter file galdef\_44\_599278pub in GALPROP), which was recently found to provide a good description of fluxes measured with the {\it Advanced Composition Explorer} for particles with nuclear charge $5 \leq Z \leq 28$ in the energy range $\sim 50$--$550$~MeV~nucleon$^{-1}$ \citep{lav13}. In Figure~\ref{fig5}, we compare the GALPROP $^{10}$Be/Be ratio and $^{10}$Be flux with the values adopted by \citet{des04}. The energy dependence of the $^{10}$Be flux results from a combination of the $^{10}$Be differential production yield and the subsequent transport of the fast ions in the ISM. The predicted decrease of the flux with decreasing energy is consistent with a recent measurement of the B flux (also a purely secondary element) in the local ISM by the Voyager 1 spacecraft \citep{cum13}. Noteworthy, another commonly-used GALPROP model that do not include the effects of diffusive reacceleration in the ISM (the plain diffusion model; parameter file galdef\_44\_999726pub) predicts an even steeper decline of the $^{10}$Be flux at low energies. 

Following \citet{des04}, we took into account both the magnetic focusing and mirroring of ions in the cloud core (using $B_{\rm ISM} = 4$~$\mu$G and $B_{\rm core} =  30$~$\mu$G), although their combine effect have little influence of the final result. We used for the core the same density profile as these authors, but we calculated the energy losses of the fast ions with the SRIM code. We find that the $^{10}$Be GCRs trapped in the core have low initial energies; it is only during the last million year before the gravitational collapse that $^{10}$Be ions of more than $10$~MeV~nucleon$^{-1}$ can be stopped in the core. 

We also took into consideration that the GCR flux at the time of the solar system formation was substantially larger than its present value. The best proxy for the GCR flux at a given time is most likely the SN rate. We note that models of Li, Be and B evolution as a function of stellar metallicity [Fe/H] depend only weakly on the time evolution of the GCR flux, because both the production of light elements and that of Fe vary with the SN rate \citep[see][]{pra12}. However, in the models of chemical evolution of the solar neighbourhood of \citet{pra08} and \citet{zhu08}, the local SN rate 4.57~Gyr ago was higher than now by factors of 1.4 and 1.6, respectively. We thus adopted an enhancement factor for the total GCR flux of 1.5, and multiplied it by a factor of 0.83 to take account of the lower $^{10}$Be/H ratio in the GCRs in the past \citep{des04}. 

Considering a starting cloud core free of $^{10}$Be, the maximum of the $^{10}{\rm Be}/ ^{9}{\rm Be}$ ratio is reached after a typical time of 4~Myr (corresponding to two times the $^{10}$Be mean lifetime). For time greater than 4~Myr, the $^{10}{\rm Be}/^{9}{\rm Be}$ ratio drops due to dilution in increasing amounts of $^{9}$Be from the cloud collapse. Under the prescriptions detailed above, we find that the trapping of GCR $^{10}$Be by the protosolar cloud core results in a maximum $^{10}{\rm Be}/^{9}{\rm Be}$ ratio of $7.7 \times 10^{-6}$, which is $\sim 130$ times less than the result of \citet{des04}. Noteworthy, our calculated $^{10}{\rm Be}/^{9}{\rm Be}$ ratio should be considered as an upper limit, because we assumed the $^{10}$Be flux to be constant below $10$~MeV~nucleon$^{-1}$ ($F=3.06 \times 10^{-10}$~cm$^{-2}$~s$^{-1}$~sr$^{-1}$~(MeV/nucleon)$^{-1 }$, see Figure~\ref{fig5}).

A part of the discrepancy between our result and the work of \citet{des04} comes from the $^{10}$Be flux at low energies (the difference is by a factor of $\sim 15$ at 10~MeV~nucleon$^{-1}$). A second source of discrepancy may come from the energy loss treatment. In their paper, \citet{des04} indicate that they considered an effective range of $0.003$~g~cm$^{-2}$ to slow down $^{10}$Be ions from 10 to 1~MeV~nucleon$^{-1}$ (cutoff energy at which they assumed the ions to be trapped). Using SRIM tabulations, it appears that the range of a $^{10}$Be of 10~MeV~nucleon$^{-1}$ in a gas of solar composition is $0.036$~g~cm$^{-2}$ (the later value exhibits a negligible variation when considering a lower cutoff at 1~MeV~nucleon$^{-1}$). As a result, much less $^{10}$Be ions are stopped than expected by \citet{des04} and the resulting $^{10}{\rm Be}/ ^{9}{\rm Be}$ ratio in the cloud core remains far below the value recorded in FUN-CAIs.

\subsubsection{Steady-state production of $^{10}$Be by background GCRs}

$^{10}$Be is continuously produced in the Galaxy by direct spallation reactions induced by GCR protons and $\alpha$-particles off ambient $^{12}$C, $^{14}$N, and $^{16}$O nuclei (of abundances in the solar system composition $x_{\rm ^{12}C} = 2.92\times10^{-4}$, $x_{\rm ^{14}N} = 7.40\times10^{-5}$, and $x_{\rm ^{16}O} = 5.36\times10^{-4}$; \citealp{asp09}). The lifetime of $^{10}$Be being shorter than the characteristic timescale of variation of the GCR flux in the Galaxy, the $^{10}{\rm Be}$ abundance is expected to be in a steady state in the ISM. In the region of the protosolar molecular cloud, we have
\begin{equation} 
\frac{^{10}{\rm Be}}{^{9}{\rm Be}}=\frac{P_{^{10}{\rm Be}}\tau_{^{10}{\rm Be}}}{(^{9}{\rm Be}/H)_\odot},     \label{eqa} 
\end{equation}
where $P_{^{10}{\rm Be}}$ is the $^{10}$Be production rate per H atom 4.57~Gyr ago, $\tau_{^{10}{\rm Be}}=2.001\pm0.017$~Myr  the $^{10}$Be lifetime \citep{kor10}, and $(^{9}{\rm Be}/H)_\odot = 2.63\times10^{-11}$ the protosolar abundance of $^{9}$Be \citep{asp09}. We calculated the $^{10}$Be production rate in the thin target approximation, i.e. by simple integration over energy of the product of the GCR flux and the spallation cross-sections presented in Appendix~B. Using the CR interaction model developed in \citet{tat12}, we checked that the ionization losses of the fast protons and $\alpha$-particles can be neglected for $^{10}$Be production as long as the H column density of the irradiated cloud is $<10^{23}$~cm$^2$. 

\begin{figure}
\centering
\includegraphics[width=8.cm]{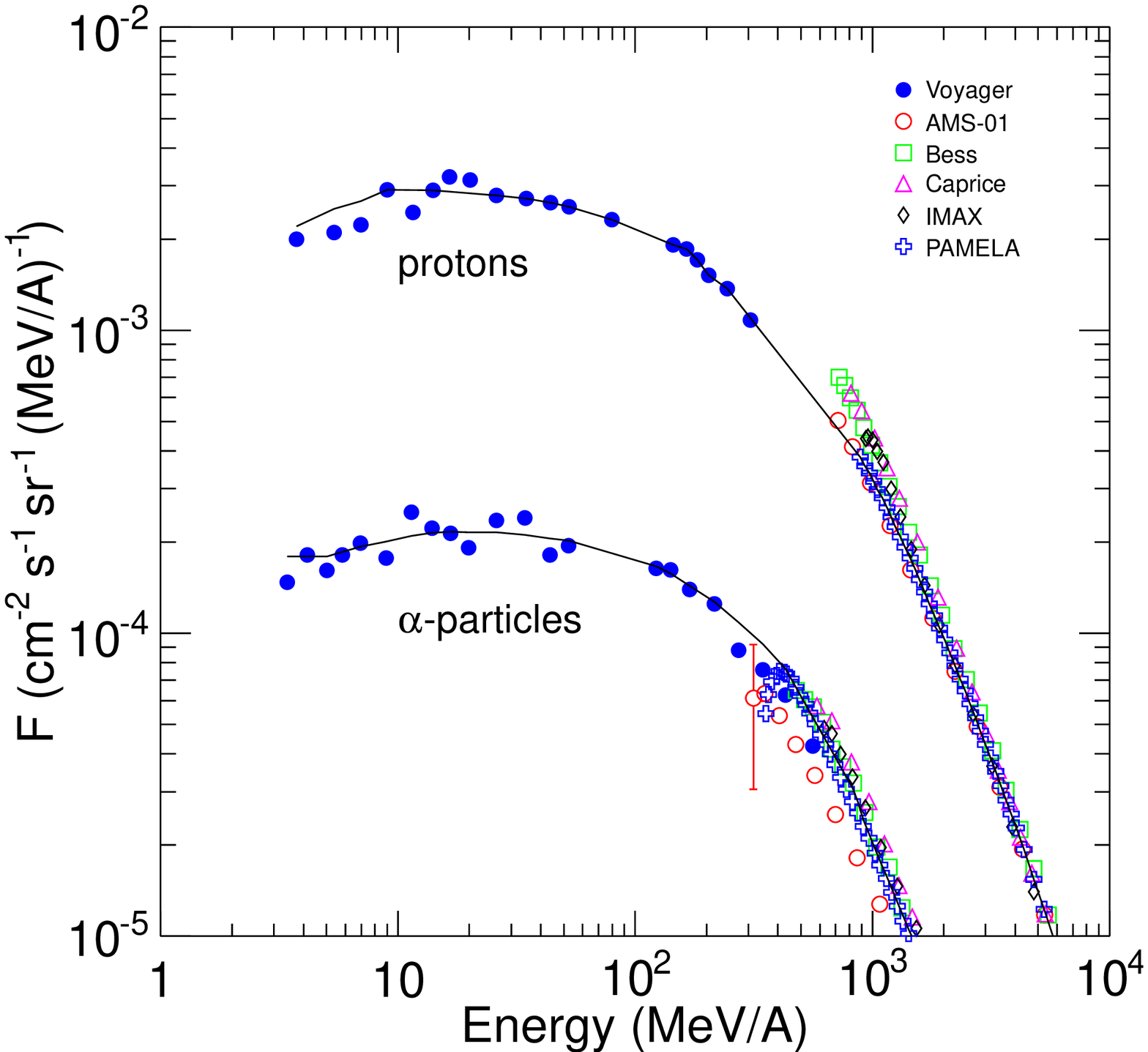}
\caption{Fluxes of GCR protons and $\alpha$ particles in the local ISM. Except for the recent data taken with the Voyager 1 spacecraft \citep{sto13}, the other observed fluxes were demodulated with the force field model of \citet{gle68} and with a solar modulation parameter depending on the epoch of observations \citep[see][and references therein]{ben13}. Solid lines (based on the Voyager and PAMELA observations only) represent the fluxes used in the present work.}
\label{gcrflux}
\end{figure}

Fluxes of GCR protons and $\alpha$-particles in the local ISM are presented in Figure~\ref{gcrflux}. Thanks to the Voyager probe that  has recently reached the edge of the solar system, these GCR spectra are now well known down to $\sim 3$~MeV~nucleon$^{-1}$ \citep{sto13}. 
We combined the Voyager data with PAMELA observations \citep{adr11} to build accurate synthetic spectra up to a few hundred GeV~nucleon$^{-1}$. 

The process of interaction of CRs with molecular clouds is not well known. \citet{ski76} pointed out that the generation of Alfven waves outside dense clouds can exclude CRs below a few hundred MeV from clouds. More recently, \citet{eve11} found that the CR density inside clouds may decrease slightly in general, and by an order of magnitude in some cases. However, we assume here that low-energy CRs can freely penetrate the clouds and that the $^{10}$Be instantaneous production rate is the same in all phases of the ISM. The $^{10}{\rm Be}/ ^{9}{\rm Be}$ ratio calculated here should thus be considered as an upper limit to the ratio produced by GCR irradiation of the presolar molecular cloud. 

With the GCR spectra determined by the Voyager and PAMELA observations, we obtain a current-epoch $^{10}$Be production rate of $1.16\times10^{-21}$~yr$^{-1}$~H$^{-1}$. The production rate is dominated by proton-induced reactions on ambient $^{12}$C and $^{16}$O, the contribution of $\alpha$ particles representing 13\% of the total. The $^{10}$Be production rate found in the present work is smaller by a factor of two than the value obtained by~\cite{des04}, $P_{^{10}{\rm Be}}=2.41\times10^{-21}$~yr$^{-1}$~H$^{-1}$. According to equation~(\ref{eqa}), we have nowadays in the local ISM $^{10}{\rm Be}/ ^{9}{\rm Be}=8.9\times10^{-5}$. With an enhancement factor of 1.5 for the GCR ion fluxes 4.57~Gyr ago (see Sect.~3.2.1), we finally get $^{10}{\rm Be}/ ^{9}{\rm Be}=1.3\times10^{-4}$ at the time of the solar system formation. Although significant, this value is substantially lower than the isotopic ratio recorded in FUN-CAIs (${\rm ^{10}Be} / {\rm ^9Be} = (2.77\pm0.24)\times 10^{-4}$ and $(3.37\pm0.20)\times 10^{-4}$ in AXCAI 2771 and KT1, respectively, the errors being at 2$\sigma$), which suggests that the presolar molecular cloud was irradiated by an additional source of CRs before its gravitational collapse. 

\subsubsection{$^{10}$Be production by CRs escaped from a nearby SNR}

\begin{figure}
\centering
\includegraphics[width=7.0cm]{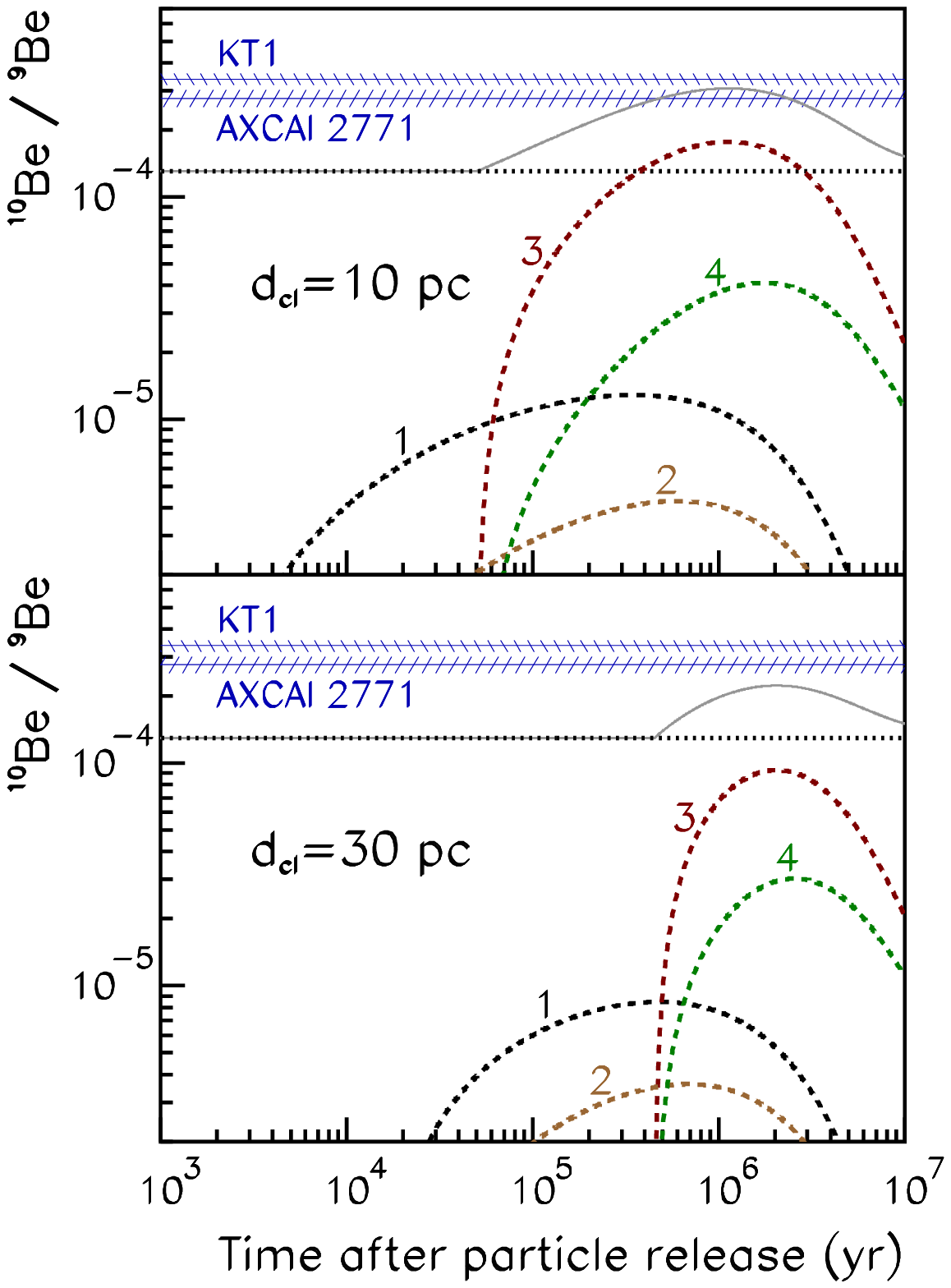}
\caption{Evolution of the $^{10}{\rm Be} / ^9{\rm Be}$ ratio in a molecular cloud irradiated by CRs escaping from an SNR, for two values of the distance between the cloud and the SNR at the end of the Sedov-Taylor phase. \textit{Upper panel}: $d_{\rm cl}=10$~pc; \textit{lower panel}: $d_{\rm cl}=30$~pc. The blue hatched areas correspond to the initial isotopic ratio found by \citet{wie12} in two FUN-CAIs: ${\rm ^{10}Be} / {\rm ^9Be} = (2.77\pm0.24)\times 10^{-4}$ in AXCAI 2771 and ${\rm ^{10}Be} / {\rm ^9Be} =(3.37\pm0.20)\times 10^{-4}$ in KT1 ($2\sigma$ errors). The horizontal dotted line shows the steady state isotopic ratio produced by background GCRs 4.57~Gyr ago ($^{10}{\rm Be}/ ^{9}{\rm Be}=1.3\times10^{-4}$). The dashed lines show additional $^{10}$Be produced by CR protons and $\alpha$ particles released from the SNR at the end of the Sedov phase. These curves were obtained for different values of the CR diffusion coefficient and H density of the intercloud medium ~-- curves 1 and 2: $D=3\times 10^{27}$~cm$^2$~s$^{-1}$, 3 and 4: $D=10^{26}$~cm$^2$~s$^{-1}$; 1 and 3: $n_{\rm H}=1$~cm$^{-3}$, 2 and 4: $n_{\rm H}=0.1$~cm$^{-3}$ (see text). The solid lines show the sum of the $^{10}{\rm Be} / ^9{\rm Be}$ ratio produced in case 3 and the steady state isotopic ratio produced by the background GCRs.}
\label{fig6}
\end{figure}

Here, we consider a model in which an additional amount of $^{10}$Be is produced in the presolar molecular cloud by spallation reactions induced by protons and $\alpha$-particles escaped from a nearby SNR (model 3 in Figure~\ref{schema10be}). This model can be applied to a massive star that evolved within its parent OB association, in the vicinity of a molecular cloud complex. The thermalization of multiple stellar winds emitted from the star cluster for million years formed a large cavity of hot gas known as a superbubble, inside which the massive star finally explodes. In this context, the SN occurs at a typical distance of a few tens of parsecs from the parent molecular cloud complex and the SNR expands in an ambient medium of density $n_{\rm H}<1$~cm$^{-3}$ \citep[see][]{tat10}.  

The highest-energy CRs accelerated at a SN blast wave continuously escape the remnant into the ISM during the Sedov-Taylor phase \citep{cap10,ell11}. However, these high-energy particles ($E \gg 1$~GeV) produce little $^{10}$Be. Spallation nucleosynthesis is mainly due to non-relativistic CRs, which remain efficiently trapped within the remnant before it enters the radiative stage. For simplicity, we assume that all non-relativistic CRs produced in the SNR during the Sedov phase are suddenly released in the ISM at the transition time $t_0=t_{\rm rad}$ (eq.~\ref{eq4}). At the time $t$ after explosion, these CRs will have diffused over a distance $R_{\rm diff}(t) \approx \sqrt{6 D (t-t_0)}$, where $D$ is the spatial diffusion coefficient of the fast particles in the vicinity of the SNR. We assume here isotropic diffusion, although CRs escaping an SNR may diffuse preferentially along regular magnetic field lines \citep{nav13}. For $t \gg t_0$, the CR density can be taken to be constant within the diffusion radius $R_{\rm diff}(t)$. Within this volume, the thin-target production rate of $^{10}$Be nuclei per ambient H atom can be written as
\begin{equation} 
P_{^{10}{\rm Be}} (t) = P_{^{10}{\rm Be}}^{\rm SNR} {R_s^3 -R_{\rm in}^3 \over [R_s+R_{\rm diff}(t)]^3}~.
\label{eq24} 
\end{equation} 
Here, $R_{\rm in}$ and $R_s$ are the inner and outer radii of the SNR shell filled with CRs at the time $t_{\rm rad}$ after explosion (see Figure~\ref{fig2}) and $P_{^{10}{\rm Be}}^{\rm SNR} = Q_{^{10}{\rm Be}} (t_{\rm rad})/N_{\rm H}^{\rm SNR}$ is the $^{10}$Be production rate in this shell ($Q_{^{10}{\rm Be}}$ is given by equation~(\ref{eq20}) and $N_{\rm H}^{\rm SNR}$ is the total number of H atoms in the shell). From the model described in Sect.~2, we find $P_{^{10}{\rm Be}}^{\rm SNR}=1.0\times10^{-19}$ and $5.9\times10^{-21}$~yr$^{-1}$~H-atom$^{-1}$ for $n_{\rm H}=1$ and $0.1$~cm$^{-3}$, respectively. 

Equation~(\ref{eq24}) assumes that the ionization losses of the escaping CRs can be neglected, such that the CR spectrum does not vary with time inside the diffusion volume, apart from the normalization. As before (Sect.~3.2.2), it is a valid approximation as long as the H column density of the irradiated cloud is $<10^{23}$~cm$^2$ \citep[see][]{tat12}. We also consider (as in Sect.~3.2.2) that the CR spectrum is not altered by the process of CR penetration into the molecular cloud. Equation~(\ref{eq24}) also assumes that the entire volume of the SNR is filled with CRs at the time $t$. 

The net rate of variation of the $^{10}{\rm Be} / ^{9}{\rm Be}$ ratio in the irradiated cloud is given by
\begin{equation} 
{d \over dt}\left({{\rm ^{10}Be} \over {\rm ^9Be}}\right)  = {P_{^{10}{\rm Be}} (t) \over x_{^9{\rm Be}}} - {1 \over \tau_{^{10}{\rm Be}}} {{\rm ^{10}Be} \over {\rm ^9Be}}~. 
\label{eq25} 
\end{equation} 
Results obtained from a numerical solution of this equation, with $P_{^{10}{\rm Be}} (t)$ calculated at each time step from equation~(\ref{eq24}), are shown in Figure~\ref{fig6} for two values of $n_{\rm H}$ and two values of the diffusion coefficient $D$. The latter parameter is not well known. With the typical mean diffusion coefficient for the propagation of GCR nuclei in the local interstellar magnetic field $B$ \citep{ber90},
\begin{equation} 
D \approx 10^{28}\beta_j \bigg({R_j \over 1{\rm~GV}}\bigg)^{0.5} \bigg({B \over 3{\rm~\mu G}}\bigg)^{-0.5}
~~{\rm cm}^2~{\rm s}^{-1},
\label{e26}
\end{equation} 
where $\beta_j=v_j/c$ and $R_j$ is the particle rigidity, one gets for 100~MeV protons: $D\approx 3\times 10^{27}$~cm$^2$~s$^{-1}$. However, the gamma-ray emission of molecular clouds near W28 is best explained with a diffusion coefficient much smaller than the average Galactic one \citep{fuj09,gab10}, and we also use $D=10^{26}$~cm$^2$~s$^{-1}$. Such a suppression of the diffusion may be the result of an increase in the level of magnetic turbulence due to CR streaming away from the SNR. 

The irradiation of a molecular cloud located at a distance $d_{\rm cl}$ from the SNR starts at the time $t_{\rm min} \approx d_{\rm cl}^2 /6D$ after the time $t_0$ of CR escape. Then, the evolution of the CR flux depends on the diffusion coefficient: if particle diffusion is still suppressed at the distance $d_{\rm cl}$ from the SNR, the flux of CRs impinging the cloud decreases relatively slowly, which enhances the $^{10}$Be production (see Figure~\ref{fig6}). As also shown in Figure~\ref{fig6}, the $^{10}{\rm Be} / ^{9}{\rm Be}$ ratio in the cloud increases with $n_{\rm H}$, which is due to the increase of the total number of CRs produced in the SNR. Noteworthy, under our assumptions, in particular as long as the thin-target approximation applies, the $^{10}{\rm Be} / ^{9}{\rm Be}$ ratio only depends on the CR flux, but not on the ambient medium density. 

Finally, we find that this model could explain the abundance of $^{10}$Be in the protosolar molecular cloud, but only for special conditions (Figure~\ref{fig6}): (i) the molecular cloud must be located close enough to the SNR, $d_{\rm cl} \lsim 10$~pc, (ii) the particle diffusion between the SNR and the cloud must be significantly suppressed with respect to the average diffusion of the GCRs, $D \ll 3 \times 10^{27}$~cm$^2$~s$^{-1}$, and (iii) the SNR must have expanded in a medium of density $n_{\rm H} \sim 1$~cm$^{-3}$. The last requirement is particularly restrictive, because the density inside a superbubble aged over 3~Myr (the minimum lifetime of a massive star) is of the order $10^{-2}$~cm$^{-3}$ \citep{mac88,tat10}. Thus, in this scenario the SN cannot result from the explosion of a massive star within its parent OB association. 

\subsubsection{$^{10}$Be production in a giant molecular cloud by CRs escaped from several SNRs within a superbubble}

Star clusters containing more than a few thousands members can give rise to several SNe exploding, with a tight space and time correlation, within the associated superbubble of hot gas. It is therefore relevant to evaluate the $^{10}$Be budget resulting from the accumulation of CRs produced by successive SNe from a large cluster. For this purpose, inspired by the model presented in \citet[][see also Young 2014]{gou09}, we use Monte Carlo simulations of the activity of massive stars ($\geq 8~M_{\odot}$) in an OB association. First, the initial mass of the stars in the parent cluster are calculated according to the Initial Mass Function (IMF) of \cite{kro93} using the mass-generating function given in~\cite{bra06}, and assuming a maximum stellar mass of 120~$M_{\odot}$. \citet{you14} considered, as initially suggested by \citet{fry07}, that single Wolf-Rayet stars do not explode as bright SNe, but rather collapse by fallback to form massive black holes. However, such an assumption is not supported by radio observations of type Ibc SNe \citep{che06}, and do not agree with the recent spectroscopic observations of SN~2013cu \citep{gal14}. Therefore, we consider here a classical scenario where all stars of initial mass greater than 8~\msun\ end their life as a SN explosion.  We verified a posteriori that the simulation results are not strongly dependent of the maximum mass of the stars exploding as a SN, as long as it is above $\sim 30~M_{\odot}$. With the IMF of \cite{kro93}, the fraction of stars between 8 and 120~\msun\ is 2.4$\times$10$^{-3}$, corresponding to $\approx$~12 and $\approx$~48 SNe for a cluster containing 5000 and 20,000 stars, respectively. 

\begin{figure}[tpb]
   \centering
   \begin{tabular}{c}
      \includegraphics[width=8.5cm]{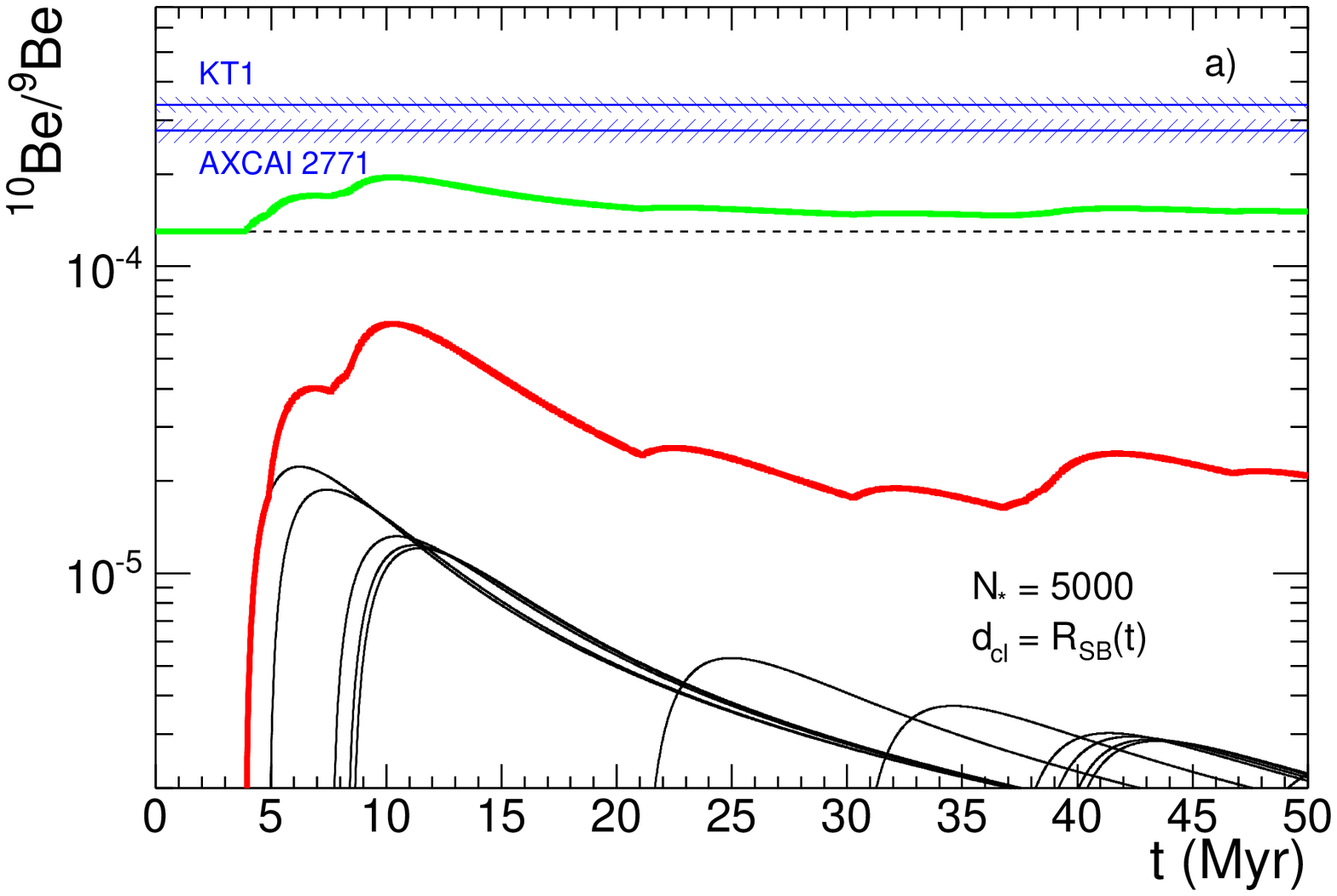} \\
      \includegraphics[width=8.5cm]{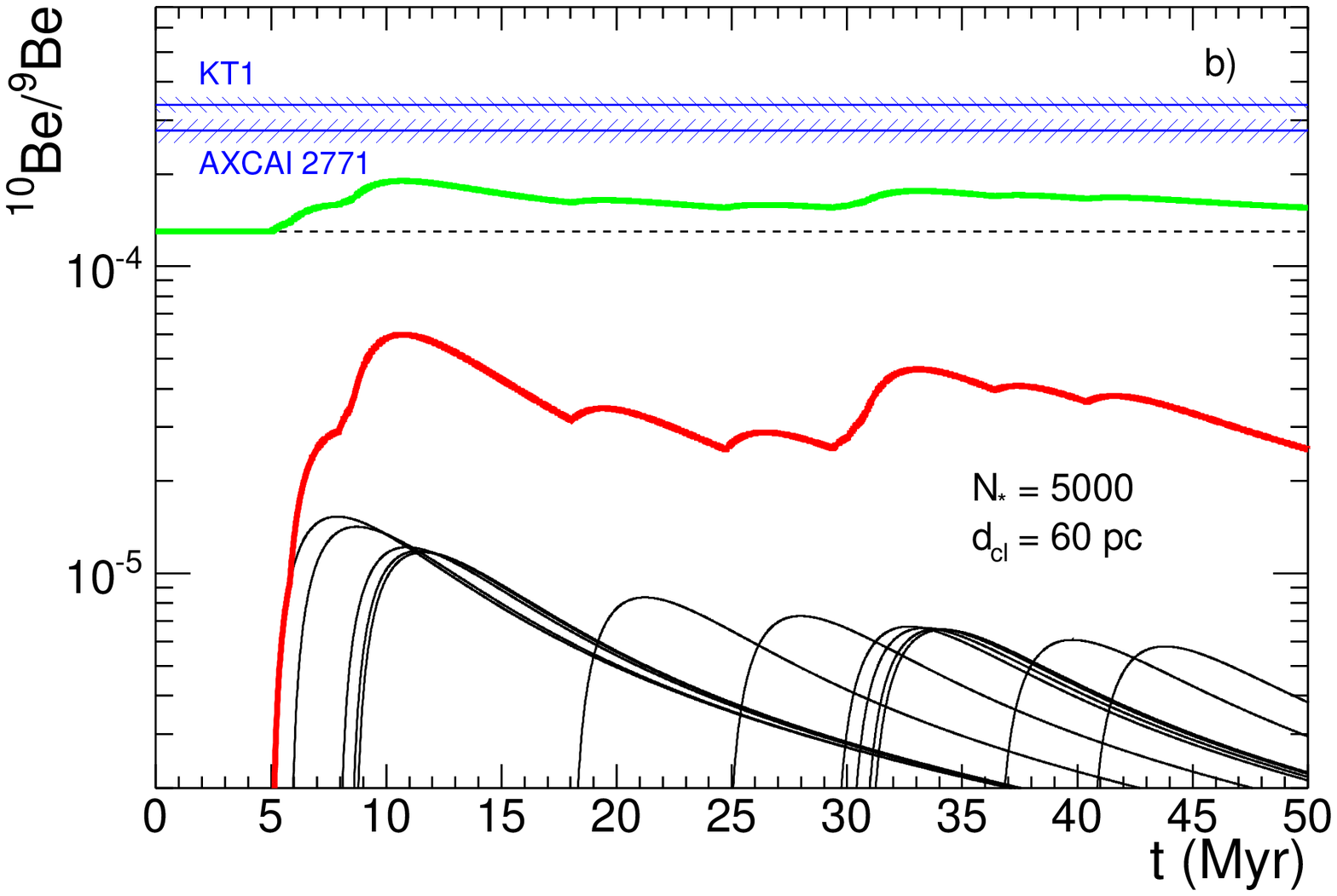}
   \end{tabular}
   \caption{Evolution of the $^{10}$Be/$^{9}$Be ratio in a molecular cloud irradiated by CRs escaped from several SNRs from a parent cluster of 5000 stars. The time origin corresponds to the birth of the star cluster. Panel ({\it a}): the molecular cloud is located at an increasing distance $R_{\rm SB}(t)$ from the SN explosions (see text); panel ({\it b}): the molecular cloud is at a fixed distance $d_{\rm cl} = 60$~pc (see text). Individual SNR contributions to the $^{10}$Be/$^{9}$Be ratio are represented by the black solid lines, while the cumulative production is shown in red. The green solid line shows the sum of the $^{10}$Be/$^{9}$Be ratio produced by CRs accelerated in the superbubble and the steady state isotopic ratio produced by the background GCRs (black dashed line). The data of \citet{wie12} are shown by the blue hatched areas, as in Figure~\ref{fig6}. The calculations assume a CR diffusion coefficient $D = 10^{26}$~cm$^{2}$ s$^{-1}$ and a H density in the molecular cloud $n_{\rm H} = 100$~cm$^{-3}$.}
   \label{f:sbb1}
\end{figure}

The explosion time ($t_i$) of each massive star is calculated using the stellar lifetimes from~\cite{sch92}. For a solar metallicity ($Z = 0.02$) the explosion times are ranging from 3 to 39~Myr for the most and less massive stars considered here. Given the number of SNe in the cluster, the density inside the superbubble $n_{\rm SB}$ and its radius $R_{\rm SB}$ are calculated at each $t_i$ using the prescriptions given in~\citet[][eqs.~1 and 3]{tat10}. The last step of the simulation calculates the production of $^{10}$Be in the nearby molecular cloud induced by each SNR following the approach presented in Sect 3.2.3. This is done by solving Eq.~(\ref{eq25}) where the inner radius of the SNR shell filled with CRs at the time $t_{\rm rad}$ ($R_{\rm in}$) and the $^{10}$Be production rate in this shell ($P_{^{10}{\rm Be}}^{{\rm SNR}}$) depend on the density in which evolves the SNR. These parameters are calculated for the density of the superbubble when the considered star explodes, i.e. $n_{\rm SB}(t_i)$. In order to take into account the stochastic nature of star formation our Monte Carlo model is repeated for many star clusters of the same size until a stable average trend is obtained for the time evolution of the $^{10}$Be/$^{9}$Be isotopic ratio.

Figure~\ref{f:sbb1} shows the result of a single Monte Carlo realization for a cluster of 5000 stars when the associated superbubble expands in a molecular cloud of H density $n_{\rm H} = 100$~cm$^{-3}$. Two scenarios are presented. The first one (Figure~\ref{f:sbb1}a) considers that the superbubble creates an expanding spherical cavity in its parent molecular cloud so that the molecular cloud is located at the distance $R_{\rm SB}(t)$ from the exploding SNe assumed to be at the center of the superbubble. The contribution of individual SNR to the production of $^{10}$Be decreases with increasing explosion times due to the dilution effect caused by the superbubble expansion (see Eq.~(\ref{eq24})). In this model, the obtained $^{10}$Be/$^{9}$Be ratio should be considered as an upper limit since we neglected the ongoing superbubble expansion during the time it takes for CRs to diffuse to the molecular cloud. As an example, a 8~\msun\ star explodes as a SN 39~Myr after the birth of the cluster and at that time the superbubble has a radius $R_{\rm SB}$ = 165~pc. The time needed for the CRs to reach such a distance is 14~Myr (for $D = 10^{26}$~cm$^{2}$~s$^{-1}$) during which the superbubble has further expanded to $R_{\rm SB} = 198$~pc. In the end, the CRs will reach the molecular cloud at a distance of $\approx$~225~pc after $\approx$~24~Myr. These additional delay and dilution factor result in an effective smaller $^{10}$Be production.

In the second scenario (Figure~\ref{f:sbb1}b), we consider a geometric configuration representative of a non-spherical superbubble expanding preferentially away from the parent molecular cloud complex due to the lower density of the ambient medium in this direction (see, e.g., \cite{bur93} for the geometry of the Orion-Eradinus superbubble with respect to the Orion molecular cloud complex). In this case, we simply assume that the molecular cloud is located at a fixed distance ($d_{\rm cl} = 60$~pc) from the place of the SN explosions. For comparison with the previous scenario, the same sequence of massive stars has been generated. When the most massive star of the present Monte Carlo realization explodes ($M = 106~M_\odot$), the superbubble radius is $R_{\rm SB} = 36$~pc, which is smaller than $d_{\rm cl}$. Therefore, the contribution of this specific SNR to the $^{10}$Be/$^{9}$Be ratio comes later and is less than in the case where the molecular clould is always at the edge of the superbubble. On the contrary, as soon as $R_{\rm SB}$ is greater than $d_{\rm cl}$, the $^{10}$Be/$^{9}$Be production ratio comes earlier and is enhanced compared to the first scenario. Noteworthy, in the second scenario, the propagation time of CRs to the molecular cloud is by assumption the same for all SNe (1.8~Myr).

\begin{figure}[tpb]
   \centering
   \begin{tabular}{c}
      \includegraphics[width=8.5cm]{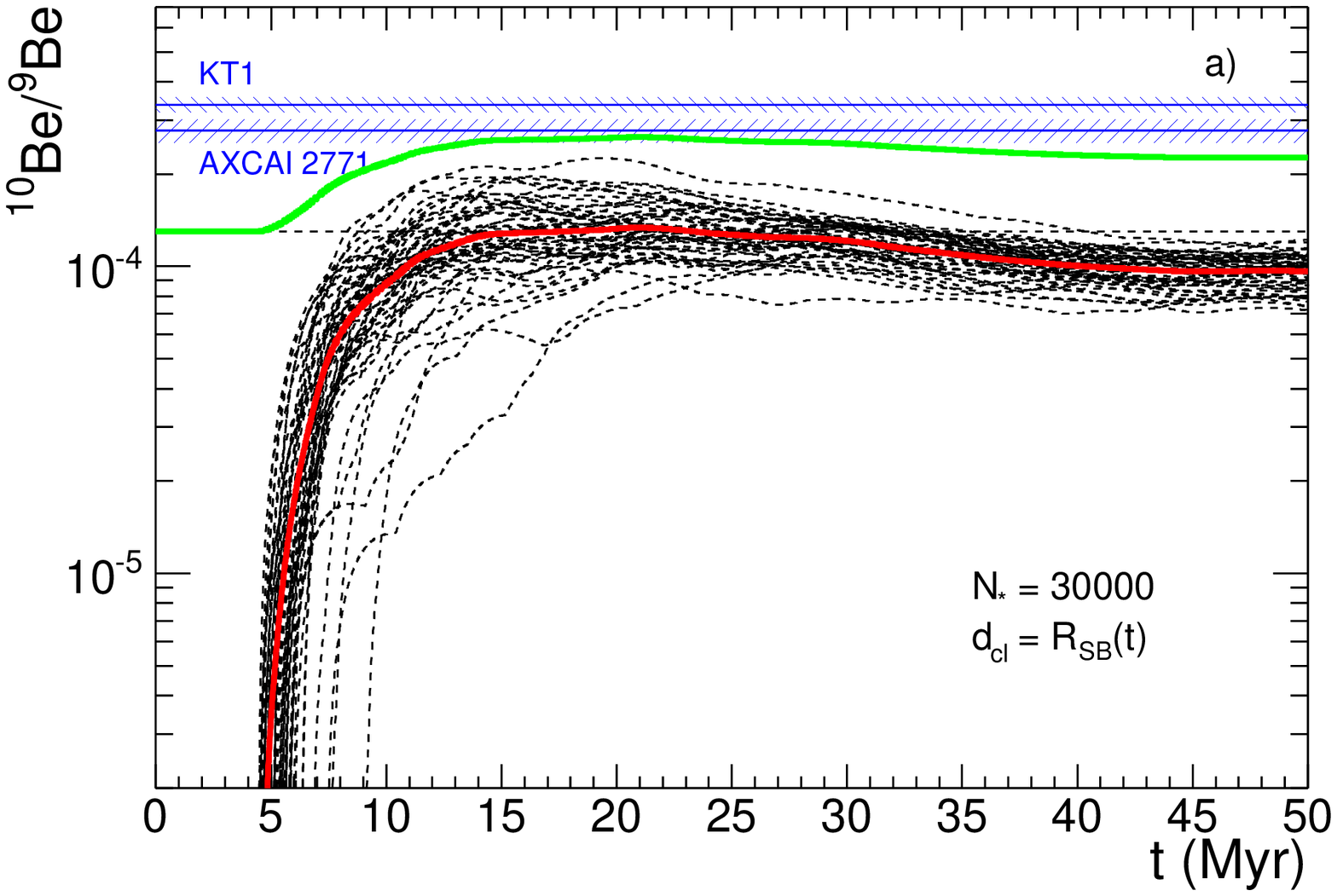} \\
      \includegraphics[width=8.5cm]{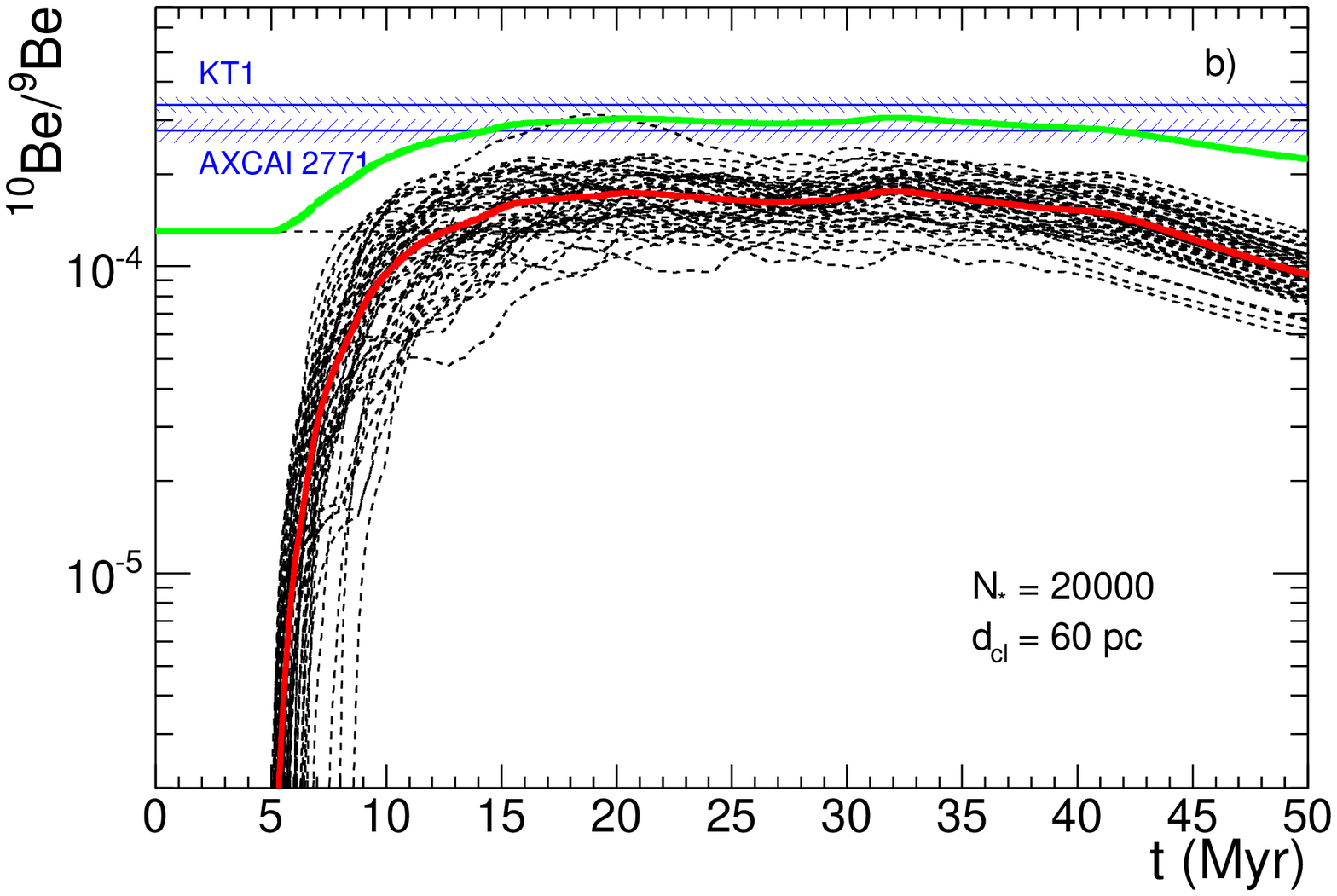}
   \end{tabular}
   \caption{Time evolution of the $^{10}$Be/$^{9}$Be isotopic ratio produced in a giant molecular cloud by a cluster of 30,000 stars with $d_{\rm cl} = R_{\rm SB}(t)$ (\textit{upper panel}) and a cluster of 20,000 stars with $d_{\rm cl} = 60$~pc (\textit{lower panel}). Each Monte-Carlo realization is represented by a black dashed line and the average of 50 realizations is shown by the red solid line. The green solid line shows the sum of the average result and the steady state $^{10}$Be/$^{9}$Be ratio produced by the background GCRs (black dashed line). The data of \citet{wie12} are shown by the blue hatched areas, as in Figure~\ref{fig6}.}
   \label{f:sbb2}
\end{figure}

We considered a whole range of cluster size, from 5000 up to several 10$^5$. For each cluster size, the Monte Carlo procedure was repeated 50 times and the mean of all realizations was calculated (red solid curves in Figures~\ref{f:sbb2}a and b). We checked that increasing the number of realizations do not significantly change the mean result. For a cluster size of 5000 stars, none of the two scenarios presented above is able to reproduce the $^{10}$Be/$^{9}$Be isotopic ratio found by~\cite{wie12} in the AXCAI 2771 and KT1 FUN-CAIs. In the scenario where the molecular gas is located at the edge of the expanding superbubble, a cluster size of $\gsim$30,000 stars is required to account for the meteoritic observations (Figure~\ref{f:sbb2}a). By contrast, a smaller cluster size of $\approx$~20,000 stars is needed in the second scenario (Figure~\ref{f:sbb2}b). In both cases, the $^{10}$Be/$^{9}$Be isotopic ratio reaches a steady state value 10~--~15~Myr after the formation of the cluster and then starts to decrease after about 40~Myr. This time interval is consistent with the lifetime of giant molecular clouds \citep[$27\pm12$~Myr;][]{mur11}, which suggests that most protostars formed in such an irradiated cloud can contain $^{10}$Be at a level close to that recorded by FUN-CAIs, whatever the time of gravitational collapse of the prestellar cores. 

These simulations thus show that the abundance of $^{10}$Be in the protosolar molecular cloud can be explained by the irradiation of a giant molecular cloud by CRs produced by $\gsim 50$ SNe exploding in a superbubble generated by a large star cluster of at least 20,000 members. The mass distribution of young clusters ($t \lsim$~10~Myr) is well described by the so-called Schechter function~\citep{sch76}: $dN_*/dM \propto M^{-\alpha} \exp(-M/M_*)$, with $M_* = 2\times10^5$~\msun\ and $\alpha = 2$ for Milky-Way-type spiral galaxies~\citep{lar09,por10}. Assuming a minimum stellar mass at 0.01~\msun\ \citep{bra06}, the IMF of~\cite{kro93} gives a mean stellar mass of 0.39~$M_\odot$, corresponding to a total mass of $M \approx 8\times$10$^3$~\msun\ for a cluster of 20,000 stars. With a minimum cluster mass of 100~\msun\ \citep{lar09}, the proportion of clusters more massive than $8\times$10$^3$~\msun\ is of the order of 1\%. About a dozen of clusters with masses in the range 10$^4$ -- 10$^5$~\msun\ are known in our galaxy~\citep{fig08}.

\subsubsection{$^{10}$Be production in a molecular cloud impacted by an SNR}

Here, we discuss an alternative scenario where a molecular cloud is directly impacted by a SNR from a massive star that escaped from its parent cluster. Due to dynamical interactions with other cluster stars, nearly half of O-type stars in the Galaxy acquire velocities exceeding the escape velocity from their parent OB association \citep{sto91}. We consider in this context two phases of $^{10}$Be production: during the Sedov-Taylor stage when trapped CRs interact with shocked molecular gas (Figure~\ref{schema10be}b), and then during the radiative stage when freshly accelerated CRs diffuse both upstream and downstream the shock (Figure~\ref{schema10be}c). 

In Section~2, for the sake of generality we have considered the case of an SNR expanding into an homogeneous CSM of constant density $n_{\rm H}$ and found that for $n_{\rm H} = 100$~cm$^{-3}$ (a typical density for a molecular cloud), the shocked gas inside the remnant contains $^{10}{\rm Be} / ^{9}{\rm Be} \sim 2.5\times 10^{-3}$ at the end of the adiabatic (Sedov) phase, which is nearly an order of magnitude higher than the $^{10}$Be abundance of the protosolar molecular cloud inferred by \citet{wie12}. In the more realistic case where only part of the forward shock hits a molecular cloud (as represented in Figure~\ref{schema10be}b), the $^{10}{\rm Be} / ^{9}{\rm Be}$ ratio in shocked molecular gas is expected to be of the same order of magnitude, because both the $^{10}$Be production and the number of collected $^9$Be atoms scale as the area of the contact surface between the shock and the cloud (see Section~2). 

The high $^{10}{\rm Be} / ^{9}{\rm Be}$ ratio found above assumes that the SNR interacts with a molecular cloud from the beginning of the Sedov phase. The $^{10}$Be abundance in the shocked gas would obviously be lower if the SNR initially expands into an intercloud medium of lower density, typically $n_{\rm H} \sim 1$~cm$^{-3}$, and then hits a molecular cloud only after some time $t_{\rm col} > t_{\rm ST}$. An upper limit on the time interval $\Delta t = t_{\rm rad} - t_{\rm col}$ can be obtained from the calculated $^{10}$Be production rate in the SNR shell at the time $t_{\rm rad}$: $P_{^{10}{\rm Be}}^{\rm SNR}=2.9\times10^{-17}$~yr$^{-1}$~H-atom$^{-1}$ for $n_{\rm H} = 100$~cm$^{-3}$. Assuming that the $^{10}$Be production rate remains approximately constant over $\Delta t$, we get
\begin{equation} 
\Delta t < {{\rm ^{10}Be} \over {\rm ^9Be}}{x_{^9{\rm Be}} \over P_{^{10}{\rm Be}}^{\rm SNR}}=270~{\rm yr}~.
\label{eq27} 
\end{equation} 
Here, we have taken ${\rm ^{10}Be} / {\rm ^9Be} = 3\times 10^{-4}$. In comparison, the Sedov-Taylor stage lasts $\sim 3\times10^4$~yr for $n_{\rm H} = 1$~cm$^{-3}$ and $\sim 2800$~yr for $n_{\rm H} = 100$~cm$^{-3}$ (eq.~\ref{eq4}). 

During the radiative stage of the SNR evolution, until the gravitationnal collapse of the protosolar molecular cloud, an additional amount of $^{10}$Be is synthesized in the cloud by CRs diffusing both upstream and downstream the forward shock. This quantity can be estimated using the simple formalism given in Section~3.2.3, with $d_{\rm cl}=0$~pc. If we adopt $n_{\rm H} = 100$~cm$^{-3}$ to estimate the number of CRs accelerated during the Sedov-Taylor phase, together with $D=10^{26}$~cm$^2$~s$^{-1}$, we find that the $^{10}{\rm Be} / ^{9}{\rm Be}$  ratio produced during the radiative stage reaches a maximum of $2.2\times 10^{-3}$ about $0.3$~Myr after the start of this phase. This value is again much higher than the estimated protosolar abundance of $^{10}$Be. But with $n_{\rm H} = 1$~cm$^{-3}$, the maximum $^{10}{\rm Be} / ^{9}{\rm Be}$ ratio is reduced to $2.1\times 10^{-4}$, which is in better agreement with the meteoritic data, taking into account the baseline ratio of $\lsim 1.3\times 10^{-4}$ produced through irradiation of the cloud by background GCRs (Sect.~3.2.2). 

For $n_{\rm H} = 1$~cm$^{-3}$, the maximum value of the $^{10}{\rm Be} / ^{9}{\rm Be}$ ratio in the molecular gas upstream the shock is reached $\sim 1$~Myr after the start of the radiative stage of the SNR evolution. Assuming that the SLR $^{41}$Ca ($t_{1/2}=0.102$~Myr) was synthesized in the SN or the progenitor star prior to explosion (see Sect.~4 below), such a delay may be too long to allow for a significant seeding of the early solar system by this radioactivity. If the gravitationnal collapse of the protosolar cloud core started $\sim 0.1$~--~$0.3$~Myr after the SN explosion, the additional amount of $^{10}{\rm Be} / ^{9}{\rm Be}$ produced by CRs escaping at the end of the Sedov phase was between $1.0 \times 10^{-4}$ and $1.7\times 10^{-4}$. 

In summary, when a molecular cloud is impacted by a young SNR, the production of $^{10}$Be inside the remnant by trapped CRs propagating in shocked molecular material should rapidly exceed the protosolar $^{10}$Be abundance recorded by  FUN-CAIs. On the other hand, if the SNR expands during most of the Sedov phase in an intercloud medium of density of $\sim 1$~H-atom~cm$^{-3}$ and interacts with the molecular cloud only during the radiative stage, the number of $^{10}$Be nuclei produced in the cloud by freshly accelerated CRs escaped from the remnant can explain the meteoritic data. 

\section{Discussion}

Of the five models considered above, two appear to be able to explain the Be isotopic ratio deduced from FUN-CAIs: (i) the irradiation of a giant molecular cloud by CRs produced by a large number of SNe exploding in a superbubble generated by a massive OB association and (ii) the impact of the presolar molecular cloud by an isolated SNR in the radiative stage of its evolution. A key difference between these two models is that in the second case, it is likely that the collapse of the presolar cloud core was triggered by a SN shock, as first suggested by \citet{cam77}. We now discuss these two models in the context of broader scenarios that have been proposed to explain the presence of other SLRs in the early solar system.

\citet{jur13} argue that the initial ratio $^{26}{\rm Al} / ^{27}{\rm Al} \sim 5 \times 10^{-5}$ of the solar system is a common feature in star-forming regions. Their conclusion is mainly based on the observation of a large variation of the ${\rm Fe} / {\rm Al}$ abundance ratio at the surface of white dwarves, which they attribute to a pollution of the star's atmosphere by accretion of differentiated asteroids. In their model, the heat source for igneous differentiation of extrasolar asteroids is from radioactive decay of $^{26}{\rm Al}$ and an initial isotopic ratio of $^{26}{\rm Al} / ^{27}{\rm Al} \ge 3 \times 10^{-5}$ is required to melt these bodies. The conclusion of these authors challenges most models for the origin of this SLR in the early solar system, because these models generally assume that the formation of the solar system took place in an unusual context for star formation in the Galaxy, e.g. in the vicinity of an AGB or super-AGB star \citep{was94,lug12}, near a runaway Wolf-Rayet star \citep{tat10}, or in a dense collected shell around a specific massive star (of initial mass $M>30~M_\odot$) on the main sequence \citep[][]{gou12}. 

By contrast, recent works have explored a more generic solution for the origin of the SLRs, where the canonical abondances of these species are the natural consequence of chemical self-pollution of giant molecular clouds  \citep{gou09,vas13,you14,sah14}. In these models, the concentrations of SLRs produced in massive stars and SNe are higher in molecular cloud complexes than the background levels of the Galaxy, because the stellar winds and SN ejecta from OB associations preferentially enrich their parental molecular clouds. Numerical simulations showed that the abundances of $^{26}{\rm Al}$, $^{36}{\rm Cl}$, and $^{41}{\rm Ca}$ in most star-forming regions could be comparable to those in the early solar system \citep{vas13,you14,sah14}. However, the process of injection of hot stellar debris into cold molecular cloud cores is not elucidate in these models \citep[see][]{tat10}. Moreover, the relatively low abundances of $^{53}{\rm Mn}$ and $^{60}{\rm Fe}$ compared to $^{26}{\rm Al}$ in the solar protoplanetary disk is difficult to explain in this context, given that $^{53}{\rm Mn}$ and $^{60}{\rm Fe}$ are substantially produced in SNe \citep[see, e.g.,][]{vas13,sah14}. 

The present work shows that all molecular clouds in the Galaxy should contain $^{10}{\rm Be}$ at a minimum level of $^{10}{\rm Be}/ ^{9}{\rm Be} \sim 10^{-4}$ due to continuous irradiation of the clouds by GCR protons and $\alpha$-particles (Sect.~3.2.2), but also that only a few giant molecular clouds may be enriched at the level of $^{10}{\rm Be}/ ^{9}{\rm Be} \sim 3\times 10^{-4}$ recorded by FUN-CAIs, as a result of an additional, local production of CRs by nearby SNe. In Section~3.2.4, we found that more than $\sim 50$ SNe exploding in a superbubble generated by a star cluster of at least 20,000 members are needed to reach the $^{10}{\rm Be}$ enrichment measured in FUN-CAIs, and that the proportion of such large clusters in the Galaxy is of the order of a percent. Thus, the abundance of $^{10}{\rm Be}$ in the early solar system cannot be explained as a generic result of the chemical evolution of giant molecular clouds, contrary to what \citet{jur13} advocated for $^{26}{\rm Al}$. We note, in particular, that in the model with clusters of 5000 stars considered by \citet{you14}, it is not possible to produce enough $^{10}{\rm Be}$ to explain the Be isotopic ratio recorded in FUN-CAIs (see Figure~\ref{f:sbb1}).

The alternative model which could explain the level of $^{10}{\rm Be}$ found in FUN-CAIs is reminiscent of the scenario we put forward in \citet{tat10}. In this paper, we proposed that $^{26}$Al, as well as $^{41}$Ca and perhaps $^{36}$Cl originated in a massive star (of mass $M \geq 25~M_\odot$) that escaped from its parent OB association and interacted with a neighboring molecular cloud complex\footnote{The abundance of $^{36}$Cl in the early solar system is controversial. In their review paper, \citet{dau11} give $^{36}{\rm Cl} / ^{35}{\rm Cl}> 1.5 \times 10^{-5}$, based on measurements using the minor branch (1.9\%) of $^{36}$Cl decay to $^{36}$S \citep[e.g.][]{jac11}. But recent measurements using the major branch, $^{36}$Cl($\beta^-$)$^{36}$Ar (98.1\%), suggest that the concentration of $^{36}$Cl in the early solar system was much lower \citep{tur13}, at a level possibly consistent with an origin in a massive star. According to \citet{arn06}, Wolf-Rayet winds can carry enough $^{36}$Cl to account for an initial $^{36}{\rm Cl} / ^{35}{\rm Cl}$ ratio of the order of $10^{-6}$.}. In this model, $^{26}$Al, $^{41}$Ca and $^{36}$Cl were synthesized and subsequently expelled into the CSM during the Wolf-Rayet phase that preceeds the explosion of the star. These ejecta were efficiently mixed with ambient molecular gas within the hydrodynamically instable bow shock resulting from the supersonic motion of the runaway star. With the nucleosynthesis yields of Wolf-Rayet stars given by \citet{pal05} and \citet{arn06}, we found that a total mass of molecular gas as high as $\sim 2\times 10^4~M_\odot$ could have been contaminated by these SLRs at canonical abundances. The final explosion of the massive star as a SN then triggered the formation of many new stars from the SLR-enriched shocked gas, including the sun. 

The present paper shows that the $^{10}{\rm Be} / ^{9}{\rm Be}$ ratio recorded by FUN-CAIs can be explained by the impact of the presolar molecular cloud by an isolated SNR, which had to evolve in a medium of about $1$~H-atom~cm$^{-3}$ during the Sedov phase to produce enough CRs. This density is higher than that of the hot ISM, which makes it plausible that the SN resulted from the explosion of a massive star that escaped from its parent cluster and associated superbubble. But we also found above that the SN blast wave is unlikely to have impacted the presolar cloud before the end of the Sedov phase, because otherwise it would overproduce $^{10}$Be. This implies that our previous model for the origin of the other three SLRs must be revised. 

Detailed hydrodynamic simulations show that a low-speed radiative shock ($V_s \leq 70$~km~s$^{-1}$) impacting a dense molecular cloud core can trigger its self-gravitational collapse \citep{bos10,bos13}. These simulations also show that a simultaneous injection of shock wave material into the core is possible under certain conditions, but with a low injection efficiency of the order of $10^{-2}$. However, the low initial ratio of $^{60}{\rm Fe} / ^{56}{\rm Fe}$ found by \citet{tan12} from analyses of whole rocks and constituents of various meteorites argues against a contamination of the protosolar molecular cloud by SN ejecta. It appears more likely that the presolar molecular cloud was contaminated before the SN explosion by the $^{26}$Al-rich, $^{60}$Fe-poor wind of a progenitor Wolf-Rayet star \citep{arn97,tat10}. But further work is needed to study the mixing efficiency of Wolf-Rayet winds into a molecular cloud. 

The scenario sketched above has the potential to explain the presence in the early solar system of the four SLRs having the shortest half-lives\footnote{A hint for the presence of $^7$Be ($t_{1/2}=53.2$~days) in the early solar system was reported by \citet{cha06}, but additional analyses of meteoritic samples have not confirmed this result \citep[see, e.g.,][]{liu10}.}: $^{41}$Ca, $^{36}$Cl, $^{26}$Al, and $^{10}$Be. SLRs with longer half-lives such as $^{60}$Fe and $^{53}$Mn may have been inherited from the average interstellar medium \citep{dau11,tan12}. Noteworthy, models of enrichment of the protosolar molecular cloud by the winds of an AGB or super-AGB star \citep{was94,lug12}, or by the winds of a massive star on the main sequence \citep{gou12} do not explain the protosolar abundance of $^{10}$Be. 

FUN-CAIs are characterized by low inferred initial abundance of $^{26}$Al, with $^{26}{\rm Al} / ^{27}{\rm Al} < 5 \times 10^{-6}$, whereas the canonical  $^{26}{\rm Al}/ ^{27}{\rm Al}$ ratio in bulk CV CAIs amounts to $(5.252 \pm 0.019)\times 10^{-5}$ \citep{lar11}. Using $^{182}$Hf--$^{182}$W age dating of a newly discovered FUN-CAI from the Allende meteorite, \citet{hol13} recently showed that the low $^{26}$Al abundance in this object, $^{26}{\rm Al} / ^{27}{\rm Al} \sim 3 \times 10^{-6}$, is not due to a late formation occuring after that of the canonical CAIs, but most likely to a heterogeneous distribution of the $^{26}$Al carrier in the protosolar molecular cloud. The interpretation of these authors is that the core of the protosolar cloud was depleted in $^{26}$Al compared to the remaining cloud, due to incomplete mixing of stellar ejecta, and that FUN-CAIs rapidly formed after the core collapse by thermal processing of presolar dust aggregates initially contained in the innermost part of the core. 

Unlike $^{26}$Al and other SLRs originating in a stellar nucleosynthetic event, $^{10}$Be was most likely distributed homogeneously in the protosolar molecular cloud, given that it was produced by thin-target irradiation. The $^{10}$Be concentration recorded by FUN-CAIs thus probably represents a minimum level initially present in all other primitive refractory solids. Additional amounts of $^{10}$Be were subsequently incorporated in some objects following spallogenic nucleosynthesis within the early solar system. This interpretation is consistent with the available $^{10}{\rm Be} / ^{9}{\rm Be}$ data set from CAIs and refractory hibonites \citep[see][]{sri13}.

\section{Conclusions}

We have developed a detailed model for the spallation nucleosynthesis of light elements in an SNR, in which the treatment of CR acceleration at the blast wave and transport of energetic particles in the downstream plasma is in accordance with the constraints imposed by recent gamma-ray observations. The model includes a new formalism to describe the transport of hadronic CRs undergoing both adiabatic and Coulomb energy losses in the postshock plasma. In agreement with previous works \citep{par99a,par99b}, we find that synthesis of the stable isotopes of Li, Be, and B by this mechanism does not have a significant effect on Galactic chemical evolution. By contrast, the $^{10}$Be abundance produced in a SNR during the Sedov-Taylor phase can be higher than the inferred initial abundance of this radioisotope in the early solar system. 

We have then studied more deeply the origin of the $^{10}{\rm Be} / ^{9}{\rm Be}$ ratio recorded by FUN-CAIs \citep{wie12}. We first showed that $^{10}$Be was not produced in situ by energetic particle irradiation of the FUN-CAIs themselves, because (i) it would lead to an overproduction of $^6$Li in these objects, (ii) the initial $^{10}{\rm Be} / ^{9}{\rm Be}$ ratio would be about ten times higher in pyroxene than in melilite, contrary to the measurements, and (iii) the irradiated FUN-CAIs would have experienced too much heating to retain large nucleosynthetic anomalies in elements such as barium and strontium. We then concluded that the $^{10}{\rm Be} / ^{9}{\rm Be}$ ratio recorded by FUN-CAIs must reflect the level of $^{10}$Be contamination of the protosolar molecular cloud, as already suggested by \citet{wie12}. 

\citet{wie12} also suggested that the $^{10}$Be abundance in the protosolar molecular cloud was generated by enhanced trapping of GCRs \citep{des04}. However, using the GALPROP code to estimate the low-energy flux of $^{10}$Be nuclei in the current-epoch GCRs, we found that the trapping mechanism provides a negligible amount of $^{10}$Be in the molecular cloud core: the inferred $^{10}{\rm Be} / ^{9}{\rm Be}$ ratio is at least 40 times lower than the initial ratio measured in FUN-CAIs. 

Irradiation of the presolar molecular cloud by the mean population of GCRs in the Galaxy (the so-called CR sea) can lead to a more significant production of $^{10}$Be. Using fluxes of GCR protons and $\alpha$ particles recently measured by the Voyager 1 spacecraft at the edge of the solar system and assuming that low-energy CRs can freely penetrate molecular clouds \citep[see, however,][and references therein]{eve11}, we found that direct spallation reactions off ambient CNO nuclei produced in the ISM a steady-state $^{10}{\rm Be} / ^{9}{\rm Be}$ ratio of $1.3\times10^{-4}$ at the time of the solar system formation. This value is smaller by a factor of $\sim 2.3$ than the $^{10}{\rm Be} / ^{9}{\rm Be}$ ratio recorded by FUN-CAIs, which suggests that the presolar molecular cloud was irradiated by an additional source of CRs.

Motivated by recent models for the chemical self-pollution of giant molecular clouds \citep{gou09,you14}, we have studied if CRs produced by successive SNe exploding within a superbubble generated by an OB association can synthesize enough $^{10}$Be in its parental molecular cloud. We found that an enrichment of the molecular gas at the level of $^{10}{\rm Be} / ^{9}{\rm Be} \sim 3\times10^{-4}$ can be achieved only for a massive OB association containing initially at least 20,000 stars. The proportion of such large clusters in the Galaxy is of the order of 1\%. However, in this scenario it is not clear how the other SLRs, which contrary to $^{10}{\rm Be}$ are produced by stellar nucleosynthesis, could be efficiently delivered into the presolar cloud core \citep{tat10}.

We have finally studied a scenario in which the required additional CRs were accelerated in an isolated SNR near the presolar molecular cloud. We have then considered two sites of $^{10}$Be production: inside the SNR by trapped CRs interacting with shocked molecular gas during the Sedov-Taylor stage, and in a molecular cloud irradiated by CRs escaping from the remnant during the radiative phase. We found that $^{10}$Be production in shocked molecular gas inside the remnant should lead to a $^{10}{\rm Be} / ^{9}{\rm Be}$ ratio of about $(2$~--~$3)\times10^{-3}$ in this medium at the end of the Sedov phase, which significantly exceeds the isotopic ratio recorded by FUN-CAIs. On the other hand, if the SNR expands during most of the Sedov phase in an intercloud medium of $\sim 1$~H-atom~cm$^{-3}$ and interacts with a molecular cloud only during the radiative stage, the amount of $^{10}$Be produced in the cloud by CRs escaped from the remnant can explain the meteoritic data. Based on our previous work on the origin of $^{26}$Al in the protoplanetary disk \citep{tat10}, we propose that the SN resulted from the explosion of a Wolf-Rayet star that ran away from its parent OB association. Contamination of the presolar molecular cloud by a Wolf-Rayet wind may indeed explain the mean abundances of $^{26}$Al, $^{41}$Ca and perhaps $^{36}$Cl in the early solar system.

\acknowledgments{We would like to acknowledge an anonymous referee for suggestions that undoubtedly helped to improve the manuscript. We also would like to thank J\"urgen Kiener for many fruitful discussions and for providing us with the demodulated CR flux data shown in Figure~\ref{gcrflux}. We are also indebted to David Maurin for helping us with the $^{10}$Be production cross sections and Alain Boudard for providing us with the INCL4.6+ABLA07 nuclear reaction code before its official release. We finally acknowledge a valuable discussion with Stefano Gabici on the issue of CR escape from an SNR. This work received funding from the ANR grant 11-BS56-026-01 (OGRESSE).}

\appendix

\section{Cosmic-ray transport inside an SNR with adiabatic and Coulomb energy losses}

\begin{figure}
\centering
\includegraphics[width=7.5cm]{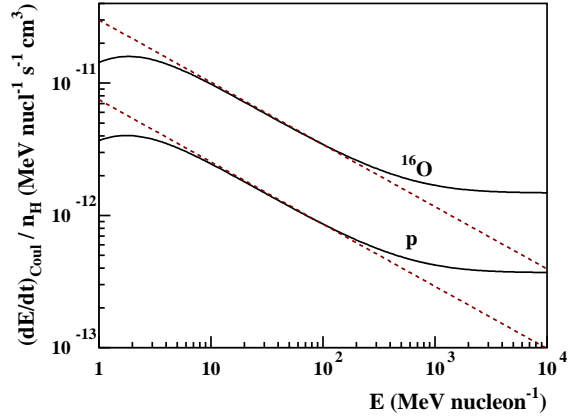}
\caption{Coulomb energy loss rate divided by the density of ambient H ions ($n_{\rm H}$) for fast protons and $^{16}$O nuclei propagating in a fully ionized plasma of electron density $n_e \cong n_{\rm H}+2 n_{\rm He} \cong 1.2n_{\rm H}$ and temperature $T=0.5$~keV. The dashed lines show a power-law fit to the energy loss rate between 10 and 200 MeV~nucleon$^{-1}$ (eq.~\ref{eqa1}).}
\label{figa1}
\end{figure}

Analytical models for the temporal evolution of accelerated particle energy distributions in SNRs have been developed to calculate the nonthermal emission of these objects \citep[e.g.,][]{stu97}, as well as light element nucleosynthesis in the early galaxy \citep{par99a,par99b}. These studies assume that the SNR interior is homogeneous and then provide the time dependence of the {\it volume-averaged} particle distribution functions. Here, we propose a simple formalism to address the transport of hadronic CRs undergoing both adiabatic and Coulomb losses in an inhomogeneous downstream plasma inside an SNR. Our approach builds on the model developed by \cite{rey98} for the synchrotron emission of SNRs. Relativistic electrons in these objects are subject to adiabatic, synchrotron and inverse Compton energy losses, and Reynolds found a simplification of the problem of electron transport through the clever use of the power-law dependence of the energy-loss rate for the latter two cooling processes, $dE/dt \propto E^2$. 

Coulomb energy loss rate of fast protons and $^{16}$O nuclei propagating in a fully ionized plasma is shown in Figure~\ref{figa1}. We used eq.~(4.22) of \cite{man94} for these calculations, except that we replaced the atomic number of the fast ions by an effective charge number calculated from eq.~(18) of \cite{kov01}. We adopted for the plasma temperature $T=0.5$~keV, which is a typical value for young and intermediate-aged SNRs. We see that between $\sim 10$ and $200$~MeV~nucleon$^{-1}$, the Coulomb energy loss rate can be well approximated by a power-law function: 
\begin{equation} 
-\left(\frac{dE}{dt} \right)_{{\rm Coul}} \cong KE^{-\beta} n_{\rm H} Z^{2} /A,      
\label{eqa1}
\end{equation} 
with $\beta=0.47$ and $K=7.5\times10^{-12}$~MeV~nucleon$^{-1}$~s$^{-1}$~cm$^3$ when the kinetic energy $E$ is expressed in MeV~nucleon$^{-1}$. Here, $Z$ and $A$ are the atomic and mass numbers of the fast ions, respectively. Inspired by the work of \cite{rey98} on electron transport in SNRs, we will adopt the above power-law fit to describe the Coulomb losses. It is a sufficient approximation for our work, because (i) light element production by CRs of energies $< 10$~MeV~nucleon$^{-1}$ is negligible and (ii) above $200$~MeV~nucleon$^{-1}$ and for reasonable values of $n_{\rm H}$, the adiabatic losses are dominant (see below).  

The adiabatic energy loss rate of {\it non-relativistic} ions transported with the downstream plasma flow from the radius $R(t)$ to the radius $R(t+dt)$ is given by
\begin{equation} 
-\left(\frac{dE}{dt} \right)_{{\rm ad}}=-\frac{2}{3} \frac{E}{\alpha \left[R(t)\right]} \frac{d\alpha \left[R(t)\right]}{dt}~,     
\label{eqa2} 
\end{equation} 
where $\alpha \left[R(t)\right]=\rho\left[R(t)\right]/\rho_d$ (simply noted $\alpha$ hereafter), $\rho_d=r\rho _{\rm CSM}$ being the mass density in the immediate postshock region, $r$ being the shock compression ratio ($r=4$ for a test-particle strong shock). During the Sedov phase, $\alpha$ can be easily calculated from the self-similar solution for the hydrodynamics (Figure~\ref{fig2}). With our approximation for the Coulomb losses, the total energy loss rate can be written as
\begin{equation} 
\frac{dE}{dt} =\left(\frac{dE}{dt} \right)_{{\rm Coul}} +\left(\frac{dE}{dt} \right)_{{\rm ad}} =\frac{2}{3} \frac{E}{\alpha } \frac{d\alpha }{dt} -KE^{-\beta } n_{\rm H} Z^{2} /A.   
\label{eqa3} 
\end{equation} 
Posing $w=E/\alpha^{2/3}$, one gets from this equation
\begin{equation} 
\frac{dw}{dt} =-Kw^{-\beta } \alpha ^{-\frac{2}{3} } n_{\rm H} Z^{2} /A,      
\label{eqa4} 
\end{equation} 
which gives after integration a relation between the kinetic energy $E$ of fast ions at the radius $R(t)$ and the energy $E_i$ of the same ions generated at the forward shock position at time $t_i<t$:
\begin{equation} 
E_{i} =\left[\left(\frac{E}{\left\{\alpha \left[R(t)\right]\right\}^{2/3} } \right)^{\beta +1} +\frac{KZ^{2} }{A} (\beta +1)\int _{t_{i} }^{t}\left\{\alpha \left[R(t')\right]\right\}^{-\frac{2}{3} (\beta +1)} n_{\rm H}^{d} \left[R(t')\right] dt' \right]^{\frac{1}{\beta +1} }~,   
\label{eqa5} 
\end{equation} 
where $n_{\rm H}^{d} \left[R(t)\right]=n_{\rm H}^{\rm CSM}r\alpha \left[R(t)\right]$ is the H number density at the radius $R(t)$. We thus have
\begin{equation} 
E_{i} =\left[\left(\frac{E}{\alpha ^{2/3} } \right)^{\beta +1} +\frac{Z^{2} n_{\rm H}^{\rm CSM} }{A} \Theta (t_{i} ,t)\right]^{\frac{1}{\beta +1} }~,      
\label{eqa6} 
\end{equation} 
with 
\begin{equation} 
\Theta (t_{i} ,t)=Kr(\beta +1)\int _{t_{i} }^{t}\left\{\alpha \left[R(t')\right]\right\}^{\frac{1}{3} (1-2\beta )} dt'~.     
\label{eqa7} 
\end{equation} 
This last equation simplifies nicely for $\beta=0.47$, since we then have in very good approximation (to better than 2\% after numerical integration):
\begin{equation}
\Theta (t_{i} ,t)=Kr(\beta +1)\left(t-t_{i} \right).      
\label{eqa8} 
\end{equation} 
Equations~(\ref{eqa6}) and (\ref{eqa8}) allow us to calculate merely the initial energy at the shock front of non-relativistic fast ions of given atomic and mass numbers. 

The temporal evolution of the differential number density of CRs is obtained by expressing the conservation of particle number:
\begin{equation} 
n_{j} \left(E,R,t\right)dE=n_{j} \left(E_{i} ,R_{s} (t_{i} ),t_{i} \right)\alpha dE_{i} .     
\label{eqa9} 
\end{equation} 
Since eq.~(\ref{eqa6}) gives
\begin{equation}
\frac{dE_{i} }{dE} =\left(\frac{E}{E_{i} } \right)^{\beta } \alpha ^{-\frac{2}{3} (\beta +1)} ,       
\label{eqa10} 
\end{equation} 
we have
\begin{equation} 
n_{j} \left(E,R,t\right)=n_{j} \left(E_{i} ,R_{s} (t_{i} ),t_{i} \right)\left(\frac{E}{E_{i} } \right)^{\beta } \alpha ^{(1-2\beta )/3} .     
\label{eqa11} 
\end{equation} 
It is noteworthy that the term $\alpha ^{(1-2\beta )/3}$ is always very close to one and can thus be neglected in this equation. But if it is the Coulomb losses that can be neglected (in practice if $n_{\rm H}^{\rm CSM} \ll 1$~cm$^{-3}$), then $\Theta=0$ and we find from eqs.~(\ref{eqa6}) and (\ref{eqa11}): $E_{i}=E\alpha^{-2/3}$ and $n_{j} \left(E,R,t\right)=n_{j} \left(E_{i} ,R_{s} (t_{i} ),t_{i} \right)\alpha ^{1/3}$. 

The above formalism applies only to non-relativistic energies, due to the approximation used for the adiabatic energy loss rate (eq.~\ref{eqa2}). In the general case, we have to use the rate of momentum loss:
\begin{equation} 
-\left(\frac{dp}{dt} \right)_{{\rm ad}} =-\frac{1}{3} \frac{p}{\alpha } \frac{d\alpha }{dt} ,       
\label{eqa12} 
\end{equation} 
which is valid whatever $p$. If the Coulomb losses are neglected, we get from this equation $p_i=p \alpha^{-1/3}$, which gives in terms of kinetic energy per nucleon, 
\begin{equation} 
E_{i} =\sqrt{\left(E^{2} +2Em_{p} c^{2} \right)\alpha ^{-2/3} +m_{p}^{2} c^{4} } -m_{p} c^{2} ,     
\label{eqa13} 
\end{equation} 
$m_p$ being the proton mass. The derivative of $E_i(E)$ is
\begin{equation} 
\frac{dE_{i} }{dE} =\frac{\alpha ^{-2/3} \left(E+m_{p} c^{2} \right)}{E_{i} +m_{p} c^{2} } ,      
\label{eqa14} 
\end{equation} 
such that we get from the equation of conservation of particle number (eq.~\ref{eqa9}):
\begin{equation} 
n_{j} \left(E,R,t\right)=n_{j} \left(E_{i} ,R_{s} (t_{i} ),t_{i} \right)\frac{\alpha ^{1/3} \left(E+m_{p} c^{2} \right)}{E_{i} +m_{p} c^{2} } .    
\label{eqa15} 
\end{equation} 
Equations~(\ref{eqa13}) and (\ref{eqa15}) provide us with a formalism correctly handling adiabatic losses whatever the energy, but not taking into account the Coulomb losses. It is therefore appropriate at sufficiently high energies for that the Coulomb losses are negligible. In practice, we use these two equations above $E_t=200$~MeV~nucleon$^{-1}$ and eqs.~(\ref{eqa6}) and (\ref{eqa11}) below this energy. 

\section{$^{10}$Be production cross sections}

\begin{figure}
\centering
\includegraphics[width=12.cm]{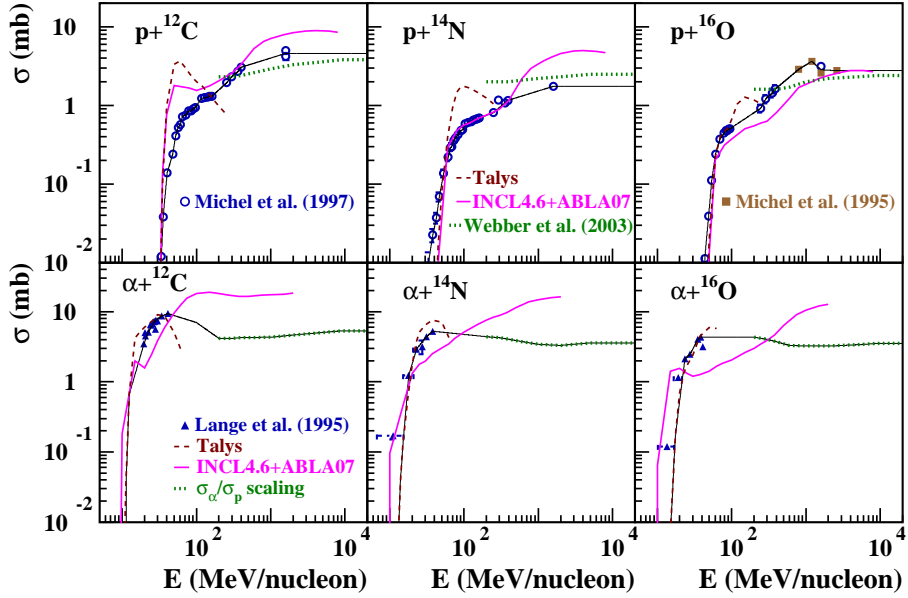}
\caption{Cross sections for the production of $^{10}$Be from proton ({\it upper panels}) and $\alpha$-particle ({\it lower panels}) interactions with $^{12}$C, $^{14}$N, and $^{16}$O. The experimental data are from \cite{mic95,mic97} for the proton reactions and \cite{lan95} for the $\alpha$ reactions. The theoretical curves were obtained with the nuclear reaction codes TALYS ({\it dashed lines}) and INCL4.6+ABLA07 ({\it thick solid lines}), the semi-empirical formula of \cite{web03} for proton reactions above $E_p=200$~MeV, and the semiempirical cross section ratio $\sigma_\alpha /\sigma_p$ of \cite{fer88} for $\alpha$ reactions above $200$~MeV~nucleon$^{-1}$. The thin solid lines show the cross section values adopted in the present work.}
\label{figb1}
\end{figure}

Total cross sections for the production of $^{10}$Be by spallation of CNO nuclei are shown in Figure~\ref{figb1}. The cross sections for the proton-induced reactions are well measured from threshold to $\sim 2$~GeV \citep[][and references therein]{mic97}. At higher energies, we took them to be energy independent. The cross sections for the $\alpha$ reactions are less well known. The only available experimental data stop at 40~MeV~nucleon$^{-1}$ \citep{lan95}. To obtain an estimate of the $\alpha$ reaction cross sections at higher energies, we first used two different nuclear reaction codes: TALYS \citep[version 1.4; ][]{kon05} and the latest version (INCL4.6) of the Li\`ege intranuclear cascade (INC) model \citep{bou13}, coupled to the ABLA07 nuclear de-excitation model \citep{kel08}. 

The TALYS computer program accounts for the major nuclear reaction models for calculations of $\gamma$-, $n$-, $p$-, $d$-, $t$-, $^3$He-, and $\alpha$-particle-induced reactions with target nuclides of mass 12 and heavier in the laboratory energy range $E_{\rm lab}=1$~keV~--~$250$~MeV. We see in Figure~\ref{figb1} that this code provides a fairly good description of the available $\alpha$ reaction data, but compares poorly with the data of \citet{mic97} for the proton-induced reactions at $E_p<250$~MeV. The code obtained by coupling INCL4.6 and ABLA07 provides a state-of-the-art description of spallation reactions induced by light particles up to $\sim 8$~GeV. We see in Figure~\ref{figb1} that the INCL4.6+ABLA07 predictions are better than those of TALYS for the proton reactions. But the predictions of the INC code for the $\alpha$ reactions are not satisfactory. They show in particular a very different behavior from one target to another around 100~MeV~nucleon$^{-1}$, which is questionable. 

In view of these results, we finally estimated the $\alpha$ reaction cross sections above 200~MeV~nucleon$^{-1}$ from the parameterisation of \citet{fer88}. These authors have developed an empirical formula for $\alpha$- to proton-induced cross section ratios, $\sigma_\alpha /\sigma_p$, based on measurements of spallation cross sections of C, O and Fe in He and H targets. We used for $\sigma_p$ the semi-empirical formulation of \citet{web03}, which gives a reasonable description of the proton data at high energies (see Figure~\ref{figb1}). The difference between the cross sections obtained from the scaling procedures of \citet{fer88} and the INCL4.6+ABLA07 predictions is up to a factor of $\sim 4$ at 2 GeV~~nucleon$^{-1}$. However, this uncertainty is not important for the present work, because most of the $^{10}$Be production in SNRs is due to proton reactions (see Figure~\ref{fig3} and Sect.~2.3.1). 

\begin{figure}
\centering
\includegraphics[width=12.cm]{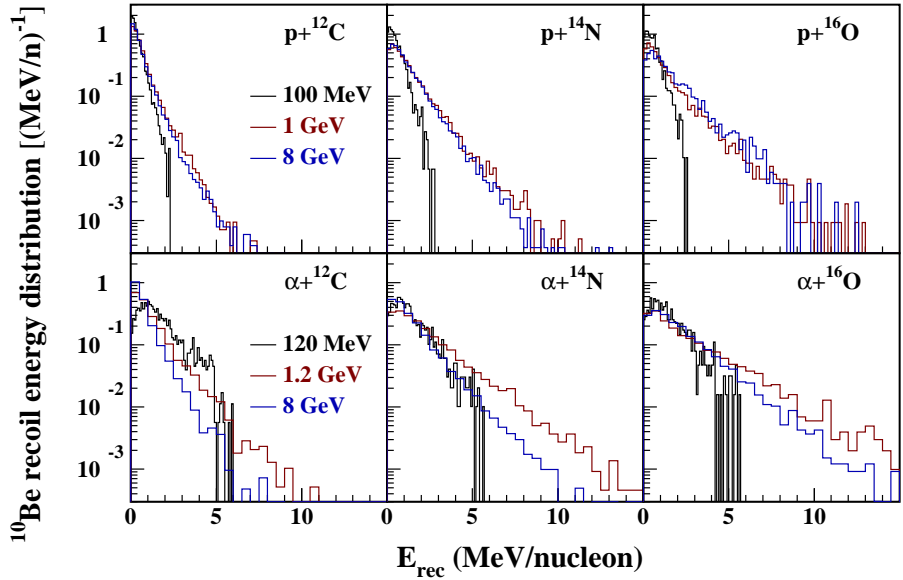}
\caption{INCL4.6+ABLA07 predictions for the recoil energy distributions of $^{10}$Be nuclei produced by proton and $\alpha$ reactions with $^{12}$C, $^{14}$N, and $^{16}$O. Proton beam energies: $E_p=0.1$, $1$, and $8$~GeV; $\alpha$ beam energies: $E_\alpha=30$, $300$, and $2000$~MeV~nucleon$^{-1}$. The distributions are normalized to unity.}
\label{figb2}
\end{figure}

As discussed in Sect.~2.3.1, some $^{10}$Be nuclei produced with a high recoil energy can escape the SNR into the ISM during the radiative phase. Calculated energy distributions of recoil $^{10}$Be ions are shown in Figure~\ref{figb2}. We see that for proton reactions, almost all $^{10}$Be nuclei are produced with recoil energy $E_{\rm rec}< 8$~MeV~nucleon$^{-1}$. For $n_{\rm H} \gsim 1$~cm$^{-3}$, these ions stop in the remnant before it becomes radiative (eq.~\ref{eq17}). The lower panels of Figure~\ref{figb2} show that the vast majority of $^{10}$Be produced in $\alpha$ reactions are trapped in the SNR as well. Thus, for $E_\alpha=300$~MeV~nucleon$^{-1}$, the calculated fractions of $^{10}$Be produced with $E_{\rm rec} > 8$~MeV~nucleon$^{-1}$ are only $\sim 0.2$\%, $\sim 1.5$\%, and $\sim 4$\%, for the spallation of $^{12}$C, $^{14}$N, and $^{16}$O, respectively.

\end{document}